\documentclass{emulateapj}

\pdfoutput=1
\usepackage{hyperref}
\usepackage{color}
\usepackage{xspace}
\usepackage{ulem}

\def\hst{\textit{HST}}

\shorttitle{SL2S: Lens models}
\shortauthors{Sonnenfeld et~al.}

\def\ucsb{1}
\def\iap{2}
\def\kipac{3}
\def\asiaa{4}
\def\oxford{5}

\def\Nsys{56}
\def\NgradeA{39}
\def\Sref#1{Section~\ref{#1}\xspace}
\def\Fref#1{Figure~\ref{#1}\xspace}
\def\Tref#1{Table~\ref{#1}\xspace}

\def\eg{{\it e.g.}\,}
\def\REin{R_{\rm Ein}}

\begin{document}

\title{The SL2S Galaxy-scale Lens Sample.  III. Lens Models, Surface Photometry and Stellar Masses for the final sample}

\author{Alessandro~Sonnenfeld\altaffilmark{\ucsb}$^{*}$}
\author{Rapha\"el~Gavazzi\altaffilmark{\iap}}
\author{Sherry~H.~Suyu\altaffilmark{\ucsb,\kipac,\asiaa}}
\author{Tommaso~Treu\altaffilmark{\ucsb}$^{\dag}$}
\author{Philip~J.~Marshall\altaffilmark{\kipac,\oxford}}
%\author{Matthew~W.~Auger\altaffilmark{\ucsb}}

% Sonnenfeld, Treu
\altaffiltext{\ucsb}{Physics Department, University of California, Santa Barbara, CA 93106, USA} 
% Gavazzi
\altaffiltext{\iap}{Institut d'Astrophysique de Paris, UMR7095 CNRS - Universit\'e
 Pierre et Marie Curie, 98bis bd Arago, 75014 Paris, France}
%Suyu:
\altaffiltext{\kipac}{Kavli Institute for Particle Astrophysics and Cosmology, Stanford University, 452 Lomita Mall, Stanford, CA 94305, USA}
\altaffiltext{\asiaa}{Institute of Astronomy and Astrophysics, Academia Sinica, P.O.~Box 23-141, Taipei 10617, Taiwan}
% Marshall:
\altaffiltext{\oxford}{Department of Physics, University of Oxford, Keble Road, Oxford, OX1 3RH, UK}

\altaffiltext{*}{{\tt sonnen@physics.ucsb.edu}}
\altaffiltext{$\dag$}{{Packard Research Fellow}}

%-------------------------------------------------------------------------------

\begin{abstract}
We present \textit{Hubble Space Telescope} (\hst) imaging data and
CFHT Near IR ground-based images for the final sample of 56 candidate
galaxy-scale lenses uncovered in the CFHT Legacy Survey as part of the
Strong Lensing in the Legacy Survey (SL2S) project. The new images are used to
perform lens modeling, measure surface photometry, and estimate
stellar masses of the deflector early-type galaxies.
Lens modeling is performed on the \hst\ images (or CFHT when \hst\ is
not available) by fitting the spatially extended light distribution of
the lensed features assuming a singular isothermal ellipsoid mass
profile and by reconstructing the intrinsic source light distribution
on a pixelized grid. Based on the analysis of systematic uncertainties
and comparison with inference based on different methods we estimate
that our Einstein Radii are accurate to $\sim 3\%$.
\hst\ imaging provides a much higher success rate in confirming gravitational lenses and measuring their Einstein radii than CFHT imaging does.
Lens modeling with ground-based images however, when successful,
yields Einstein radius measurements that are competitive with
spaced-based images.
Information from the lens models is used together with spectroscopic
information from the companion paper IV to classify the systems,
resulting in a final sample of \NgradeA\ confirmed (grade-A) lenses
and 17 promising candidates (grade-B,C). This represents an increase of half an order of
magnitude in sample size with respect to the sample of confirmed
lenses studied in papers I and II.
The Einstein radii of the confirmed lenses in our sample span the
range $5-15$ kpc and are typically larger than those of other surveys,
probing the mass in regions where the dark matter contribution is more
important. Stellar masses are in the range $10^{11}-10^{12}M_\odot$, covering the range of
massive ETGs. The redshifts of the main deflector span a range $0.3\le
zd \le0.8$, which nicely complements low-redshift samples like the
SLACS and thus provides an excellent sample for the study of the
cosmic evolution of the mass distribution of early-type galaxies over the second half of the history of the Universe.
\end{abstract}


%-------------------------------------------------------------------------------

\section{Introduction}\label{sect:intro}

Strong gravitational lensing is a powerful and consolidated technique for measuring the distribution of matter in massive galaxies at cosmological distances.
Strong lensing provides, with very few assumptions, a measurement of the projected mass of a galaxy integrated within an aperture to better than a few percent.
Early-type galaxy (ETG) lenses in particular have allowed for a number of studies covering relevant topics of cosmology such as the density profile of ETGs \citep[e.g.,][]{RKK03,Rus++03,K+T04,Bar++11a}, the value of the Hubble constant and other cosmological parameters \citep[e.g.,][]{Suy++10,Suy++13,Gav++08}, 
the abundance of mass substructure in galaxies \citep[e.g.,][]{V+K09a}, the stellar initial mass function \citep[e.g.,][]{Tre++10,Fer++10} and the shape of dark matter halos \citep[e.g.,][]{Son++12,Gri12}. 
The current number of known early-type galaxy lenses is avobe two hundred. 
While some of these lenses were serendipitous findings, most of them were discovered in the context of dedicated surveys.
The largest such survey to date is the Sloan Lens ACS (SLACS) survey \citep{Bol++04}, which provided about 80 lenses.
Although this sample has yielded interesting results on the properties of ETGs, there are many astrophysical questions that can be better answered with a larger number of strong lenses spanning a larger volume in the space of relevant physical parameters.
For instance, quantities like the dark matter fraction or the density slope of ETGs, measurable with lensing and stellar kinematics information, might be correlated with other observables such as the stellar mass or the effective radius.
Moreover, the mass structure of ETGs could be evolving in time as a result of the mass accretion history. In order to test this scenario, a statistically significant number of lenses covering a range of redshift is needed.
However, most of the galaxy-scale lenses known today are limited at a redshift $z<0.3$, corresponding to a lookback time of about 3.4~Gyr.

One of the goals of the Strong Lensing Legacy Survey (SL2S) collaboration is to extend to higher redshifts the sample of known galaxy-scale gravitational lenses.
In Papers I and II \citep{Gav++12,Ruf++11} we presented the pilot sample of 16 lenses. 
Here we extend our study to a sample of \Nsys~objects at redshifts up to $z=0.8$.
In this paper we present the lensing models of the new systems along with revisited models of the old ones.
Furthermore,
we make more conservative assumptions about the intrinsic shape of the lensed sources by reconstructing them on a pixelized grid \citep{W+D03,Suy++06,K+T04}.
In a companion paper \citep[hereafter Paper IV]{PaperIV} we include the stellar kinematic measurements and address the issue of the time evolution of the density profile of ETGs.

The goal of this paper is to present our new sample of lenses, characterize it in terms of Einstein radii and stellar masses, and to compare the effectiveness of ground-based versus space-based images for the purpose of confirming gravitational lens candidates.
This paper, the third in the series, is organized as follows. \Sref{sect:sl2s} summarizes the SL2S and the associated Canada-France-Hawaii-Telescope Legacy Survey (CFHTLS) data, the lens detection method and the sample selection.
In \Sref{sect:phot} we present all the photometric data set of the SL2S lenses, either coming from the CFHTLS parent photometry or from additional \textit{Hubble Space Telescope} (\hst) and Near infrared (IR) follow-up imaging.
In \Sref{sect:lensing} we describe the lens models of the \Nsys~systems.
In \Sref{sect:mstar} we show measurements of the stellar mass of our lenses from stellar population synthesis fitting.
We discuss and summarize our results in \Sref{sect:summary}.
Throughout this paper, magnitudes are given in the AB system.  
When computing distances, we assume a
$\Lambda$CDM cosmology with matter and dark energy density $\Omega_m=0.3$,
$\Omega_{\Lambda}=0.7$, and Hubble constant H$_0$=70 km s$^{-1}$Mpc$^{-1}$.

%-------------------------------------------------------------------------------

\section{The Strong Lensing Legacy Survey}\label{sect:sl2s}

SL2S \citep{Cab++07} is a project dedicated to finding and studying galaxy-scale and group-scale strong gravitational lenses in the Canada France Hawaii Telescope Legacy Survey (CFHTLS).
The main targets of this paper are massive red galaxies.
%SHS: I changed the wording a bit to include the group-scale SL2S lenses.
The galaxy-scale SL2S lenses are found with a procedure described in detail in Paper I \citep{Gav++12} that can be summarized as follows.
We scan the 170 square degrees of the CFHTLS with the automated software {\tt RingFinder} (Gavazzi et al., in prep.) looking for tangentially elongated blue features around red galaxies.
The lens candidates are then visually inspected and the most promising systems are followed up with \hst\ and/or spectroscopy.

Previous papers have demonstrated the success of this technique. In Paper I \citep{Gav++12}, we obtained lens models for a pilot sample of 16 lenses and in Paper II \citep{Ruf++11}, we combined this information with spectroscopic data to investigate the total mass density profile of the lens galaxies, and its evolution.
Here we complete the sample by presenting all the new systems that have been followed-up with either high-resolution imaging
or spectroscopy since the start of the campaign. We also re-analyze the pilot sample to ensure consistency. This paper is focused on the sample's photometric data and lens models, while in Paper IV we present the corresponding spectroscopic observations, model the mass density profile of our lenses, and explore the population's evolution with time.

\begin{figure}
\includegraphics[width=\columnwidth]{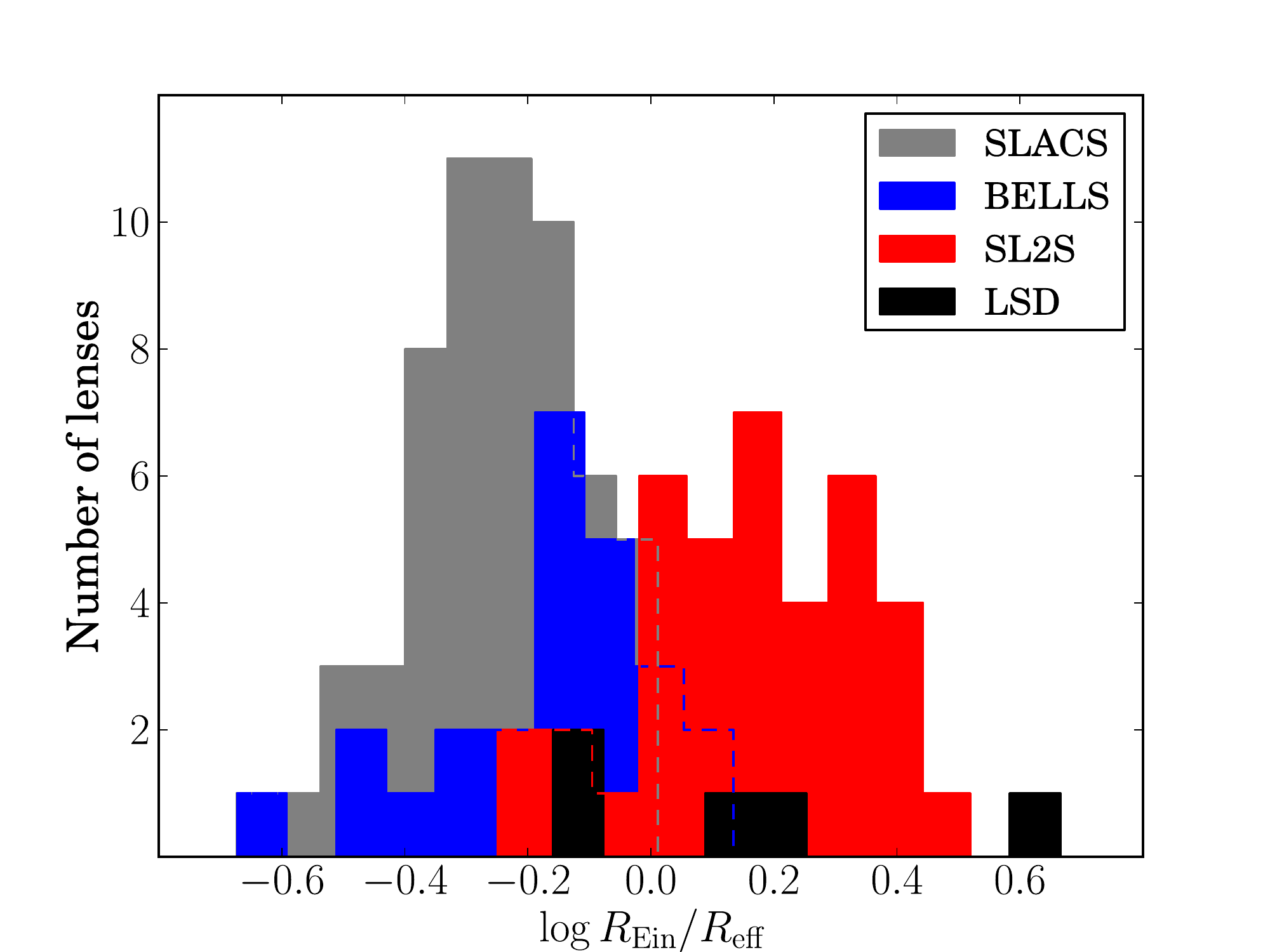}
\includegraphics[width=\columnwidth]{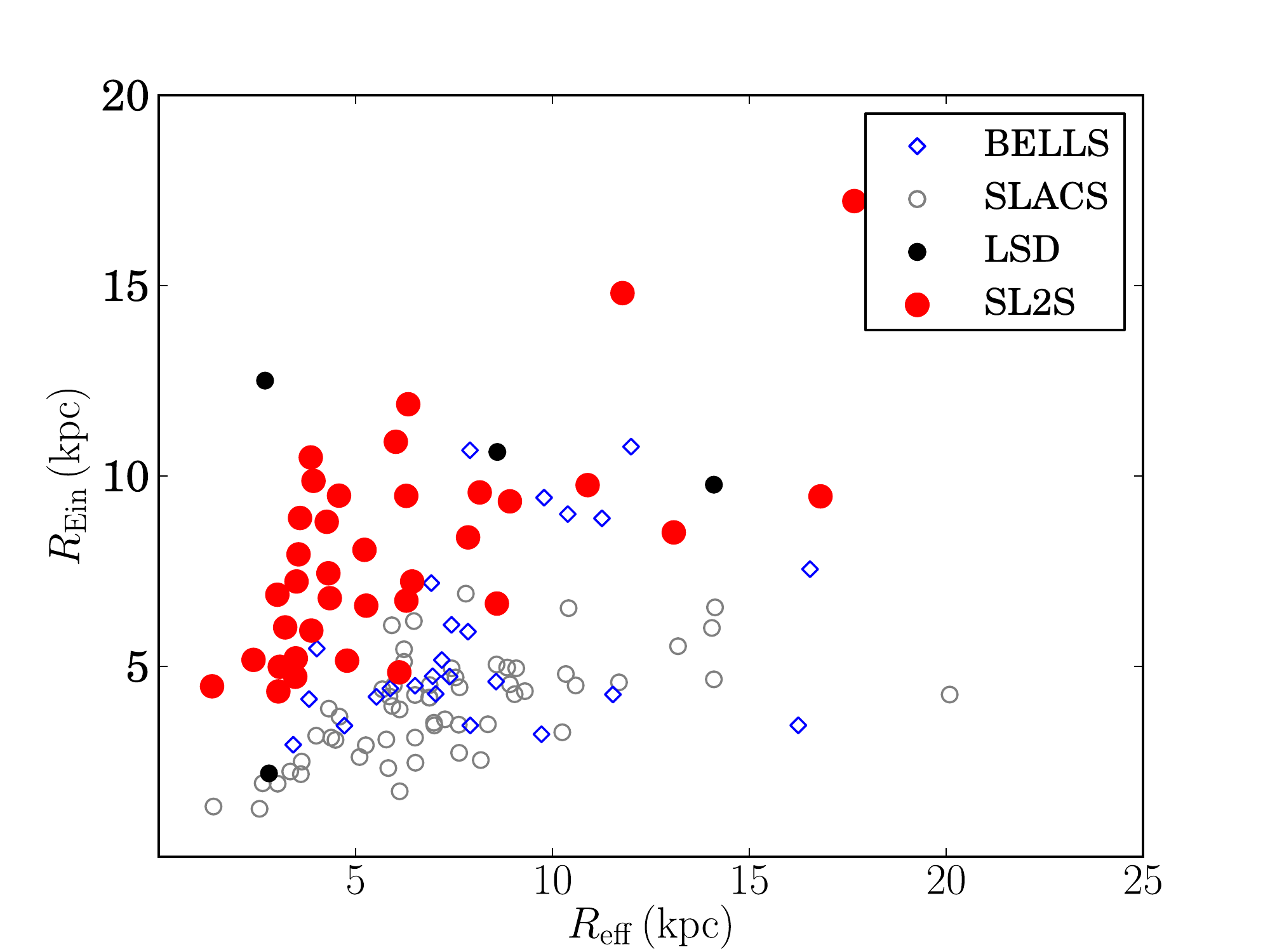}
\caption{\label{fig:reinhist} {\it Top Panel:} Distribution of Einstein radii, scaled
  by the effective radius, of lenses from SLACS \citep{Aug++10}, BELLS \citep{Bro++12}, LSD \citep{T+K04} and grade-A SL2S.\ \ \ {\it Bottom Panel:} Same samples shown in the $R_{\mathrm{Ein}}$-$R_{\mathrm{eff}}$ plane.}
\end{figure}

SL2S is by no means the only systematic survey of massive galaxy lenses: two other large strong-lens samples are those of the SLACS \citep{Bol++04} and BELLS \citep[BOSS Emission-Line Lensing Survey;][]{Bro++12} survey.
SL2S differs from SLACS and BELLS in the way lenses are found. While we look for lenses in wide-field imaging data, the SLACS and BELLS teams selected candidates by looking for spectroscopic signatures coming from two objects at different redshifts on the same line of sight in the Sloan Digital Sky Survey (SDSS) spectra.
These two different techniques correspond to differences in the population of lenses in the respective samples.
Given the relatively small fiber used in SDSS spectroscopic observations ($1\farcs5$ and $1\arcsec$ in radius, for SLACS and BELLS respectively), the spectroscopic surveys tend to select preferentially lenses with small Einstein radii, where both the arc from the lensed source and the deflector can be captured within the fiber. 
SL2S, on the other hand, finds with higher frequency lenses with Einstein radii above $1\arcsec$, since they can be more clearly resolved in ground-based images (even after the lensed sources have been deblended from the light of the central deflector).
At a given redshift, different values of the Einstein radius correspond to different physical radii at which masses can be measured with lensing.
For a quantitative estimate of the range of physical radii probed by the different surveys, we plot in \Fref{fig:reinhist} the distribution of Einstein radii and the effective radii for lenses from SL2S (determined in Sections \ref{ssec:photolenses} and \ref{sec:REin}),
% SHS: TODO: double check
 BELLS \citep{Bro++12} and SLACS \citep{Aug++10}, together with 5 lenses from the LSD study \citep{T+K04}.
The different surveys complement each other nicely, each one providing
independent information that cannot be easily gathered from the others.

%\QUERY{PJM}{What would the distribution of all galaxies look like in this figure? What story are we trying to tell regarding the selection function, and the conclusions about massive galaxies in general that we want to draw? Is it true that SL2S is ``more biased'' a sample than BELLS? Is this a question for a different paper?} 

%....................
%....................

% \subsection{Sample size and selection}\label{ssec:sample}
% PJM: Looks odd to have a subsection 2.1 and no other subsections...

In \Tref{table:census} we provide a census of SL2S targets that have been followed up so far.
The systems are graded according to their subjective
likelihood of being strong lenses:
grade A are definite lenses, B are probable lenses, C are possible
lenses or, more conservatively, systems for which the additional data set does not lead to conclusive answers about their actual strong lensing nature, and, grade X are non-lenses.
Grades for individual systems are shown in \Tref{table:lensfit} and discussed
in \Sref{ssect:lenses}.

In this paper we show detailed measurements of photometric properties, lens models and stellar masses for all grade A lenses and for all grade B and C systems with spectroscopic follow-up.
The same selection criterion is applied in Paper IV.

\begin{deluxetable}{lccccc}
 \tablecaption{ \label{table:census} Census of SL2S lenses.}
 \tablehead{
Grade & A & B & C & X & Total }
 \startdata
 With high-res imaging & 30 & 3 & 13 & 21 & 67\\ 
With spectroscopy & 36 & {\bf 15} & {\bf 2} & 5 & 58 \\ 
High-res imaging and spectroscopy & 27 & 3 & 0 & 0 & 30 \\ 
Total with follow-up & {\bf 39} & 15 & 15 & 26 & 95 \\ 

 \enddata
 \tablecomments{Number of SL2S candidates for which we obtained
   follow-up observations in each quality bin. Grade A: definite
   lenses, B: probable lenses, C: possible lenses, X:
   non-lenses. We differentiate between lenses with spectroscopic
   follow-up, high-resolution imaging follow-up or any of the two.
   In bold font we give the numbers that add up to our overall
   sample size of 56. 
}
\end{deluxetable}

%-------------------------------------------------------------------------------

\section{Photometric observations}\label{sect:phot}

All the SL2S lens candidates are first imaged by the CFHT as part of the CFHT Legacy Survey.
CFHT optical images are taken with the instrument Megacam in the $u,
g, r, i, z$ filters under excellent seeing conditions. 
The typical FWHM in the $g$ and $i$ bands is $0\farcs7$.
We refer to \citet{Gav++12} for a more detailed description of
ground-based optical observations.

The WIRCam \citep{Pug++04} mounted on the CFHT was used to get Near IR
follow-up photometry for some of the SL2S lens galaxies (Programs
11BF01, PI Gavazzi, and 07BF15 PI Soucail) or from existing surveys
like WIRDS \citep{Bie10++,Bie12++}\footnote{see also
  \url{http://terapix.iap.fr/article.php?id\_article=832}} or from an
ongoing one, called Miracles that is gathering a deep Near IR
counter-part to a subset of the CFHTLS in the W1 and W4 fields
(Arnouts et al., in prep). All these data were kindly reduced by the 
Terapix team.\footnote{\url{http://terapix.iap.fr}} 
% PJM: note that the \footnote goes *after* the punctuation. 
$Ks$ (and sometimes also $J$ and $H$) band is used for the systems listed in
\Tref{table:NIR} to estimate more accurate stellar masses (see \ref{sect:mstar}).

In addition to ground-based photometry, 33 of the \Nsys~lens systems
presented here have been observed with \hst\ as part of programs 10876, 11289 (PI Kneib) and 11588 (PI
Gavazzi), over the course of cycles 15, 16 and 17 respectively.
A summary of \hst\ observations is given in \Tref{table:HST}.
The standard data reduction described in Paper I was performed.
%The remaining ones are lenses that are either spectroscopically confirmed, by the observation of emission lines at higher redshift than the lens in correspondance to the blue lensed features, or systems for which the ground-based data shows a clear lens-like morphology can be described with a simple and robust lens model.

%Finally, we have NIR imaging with adaptive optics from the instrument
%NIRC2 on Keck Telescope. NIRC2 images of SL2S candidates have yielded
%very little information on the targets as the lensed background
%sources, selected for their blue colors, have little flux in the
%NIR. We will not discuss data coming from NIRC2 observations, but
%NIRC2 images are shown in the Appendix.

% SHS: switched order of NIR and HST table, so that it follows the order mentioned in the text.

% %%%%%%%%%%%%%%%%%%%%%%%%%%%%%%%%%%%%%
%  \begin{table}[!h]
%  \caption{Summary of NIR observations}
%  \label{table:NIR}
%  \begin{tabular}{ccccc}
%  \input{tables/rm/NIRtable.tex}
%  \end{tabular}
%  \end{table}
% %%%%%%%%%%%%%%%%%%%%%%%%%%%%%%%%%%%%%

% %%%%%%%%%%%%%%%%%%%%%%%%%%%%%%%%%%%%%
\renewcommand{\arraystretch}{1.10} 
\begin{deluxetable}{ccccc}
  \tabletypesize{\footnotesize}
  \tablecaption{\label{table:NIR} Summary of NIR observations}
  \tablehead{
  Name & Program & Filter & Exp.\ time \\
       &         &        & (s) }
  \startdata
    SL2SJ021325$-$074355 & 11BF01 & Ks & 1050 \\
SL2SJ021411$-$040502 & 07BF15 & J,Ks & 970,2480 \\
SL2SJ021737$-$051329 & 07BF15 & J,Ks & 2470,1810 \\
SL2SJ021902$-$082934 & 11BF01 & Ks & 1020 \\
SL2SJ022357$-$065142 & 11BF01 & Ks & 1000 \\
SL2SJ022511$-$045433 & WIRDS & J,H,Ks & 15720,11750,12860 \\
SL2SJ022610$-$042011 & WIRDS & J,H,Ks & 13230,10240,11300 \\
SL2SJ022648$-$040610 & WIRDS & J,H,Ks & 1800,820,1570 \\
SL2SJ023251$-$040823 & 11BF01 & Ks & 1010 \\
SL2SJ084909$-$041226 & 11BF01 & Ks & 1370 \\
SL2SJ084959$-$025142 & 11BF01 & Ks & 1580 \\
SL2SJ085826$-$014300 & 11BF01 & Ks & 1570 \\
SL2SJ090106$-$025906 & 11BF01 & Ks & 1320 \\
SL2SJ090407$-$005952 & 11BF01 & Ks & 1050 \\
SL2SJ095921+020638 & WIRDS & J,H,Ks & 7500,16270,2990 \\
SL2SJ220329+020518 & 11BF01 & Ks & 1840 \\
SL2SJ220506+014703 & MIRACLES & Ks & 1140 \\
SL2SJ220629+005728 & MIRACLES & Ks & 1340 \\
SL2SJ221326$-$000946 & 11BF01 & Ks & 1280 \\
SL2SJ221852+014038 & MIRACLES & Ks & 970 \\
SL2SJ222012+010606 & MIRACLES & Ks & 1070 \\
SL2SJ222148+011542 & MIRACLES & Ks & 250 \\

  \enddata
\end{deluxetable}
% %%%%%%%%%%%%%%%%%%%%%%%%%%%%%%%%%%%%%

% %%%%%%%%%%%%%%%%%%%%%%%%%%%%%%%%%%%%%
%  \begin{table}[!h]
%  \caption{Summary of HST observations}
%  \label{table:HST}
%  \begin{tabular}{cccccc}
%  \input{tables/rm/HSTtable.tex}
%  \end{tabular}
%  \end{table}
% %%%%%%%%%%%%%%%%%%%%%%%%%%%%%%%%%%%%%

% %%%%%%%%%%%%%%%%%%%%%%%%%%%%%%%%%%%%%
\renewcommand{\arraystretch}{1.10} 
\begin{deluxetable}{cccccc}
  \tabletypesize{\footnotesize}
  \tablecaption{\label{table:HST} Summary of \hst\ observations}
  \tablehead{
  Set & Program & Instrument & Filter & Exp.\ time \\
      &         &            &        & (s)       }
  \startdata
    (a) & 10876 & ACS   & F814W,F606W  & 800,400 \\
(b) & 11689 & WFPC2 & F606W        &    1200 \\
(c) & 11588 & WFC3  & F600LP,F475X &     720 \\
(d) & 11588 & WFC3  & F475X        &     360

  \enddata
\end{deluxetable}
% %%%%%%%%%%%%%%%%%%%%%%%%%%%%%%%%%%%%%

%-------------------------------------------------------------------------------

\subsection{Properties of lens galaxies}\label{ssec:photolenses}

We wish to measure magnitudes, colors, effective radii, ellipticities and orientations of the stellar components of our lenses.
This is done first by using the CFHT data, for all systems.
We simultaneously fit for all the above parameters to the full set of
images in the 5 optical filters, and NIR bands when available,
by using the software {\tt spasmoid}, developed by M. W. Auger and described in \citet{Ben++11}.
Results are reported in \Tref{table:cfhtphot}.
For systems with available \hst\ data we repeat the fit using \hst\ images alone.
The measured parameters are reported in \Tref{table:hstphot}.
Uncertainties on CFHT lens galaxy magnitudes are dominated by contamination from the background source and are estimated to be $0.3$ in $u$ band, $0.2$ in $g$ and $r$, $0.1$ in all redder bands, while \hst\ magnitudes have an uncertainty of 0.1. %\QUERY{RG}{Do we agree on such big errors?} \ANSWER{AS}{I would not believe more ambitious measurements given that in $g$ and $u$ the source is usually brighter than the lens}. 
Although we used the same data, some of the CFHT magnitudes previously
reported for the lenses studied in Paper I and Paper II are slightly
inconsistent with the values measured here. This difference is partly due to
a different procedure in the masking of the lensed arcs.
In Paper I and II, the lensed features were masked out automatically
by clipping all the pixels more than $4 \sigma$ above the best fit de
Vaucouleurs profile obtained by fitting the deflector light
distribution with {\tt Galfit} \citep{Pen++02,Pen++10b}, while here the masks are defined manually for every lens. We verified that this different approach is sufficient for causing the observed mismatch.
The masking procedure adopted here is more robust and therefore we consider the new magnitudes more reliable.
In addition, the measurements reported in Paper I and Paper II were allowing for different effective radii in different bands and the resulting magnitudes depend on the extrapolation of the light profile at large radii where the signal-to-noise ratio is extremely low. Here we fit for a unique effective radius in all bands, resulting in a more robust determination of relative fluxes, i.e.~colors, important for the determination of stellar masses from photometry fitting. We note that this corresponds to an assumption of negligible intrinsic color gradient in the lens galaxies. However, asserting an effective radius that is constant across bandpasses mitigates against the much larger contamination from the lensed source.

\begin{figure}
\includegraphics[width=\columnwidth]{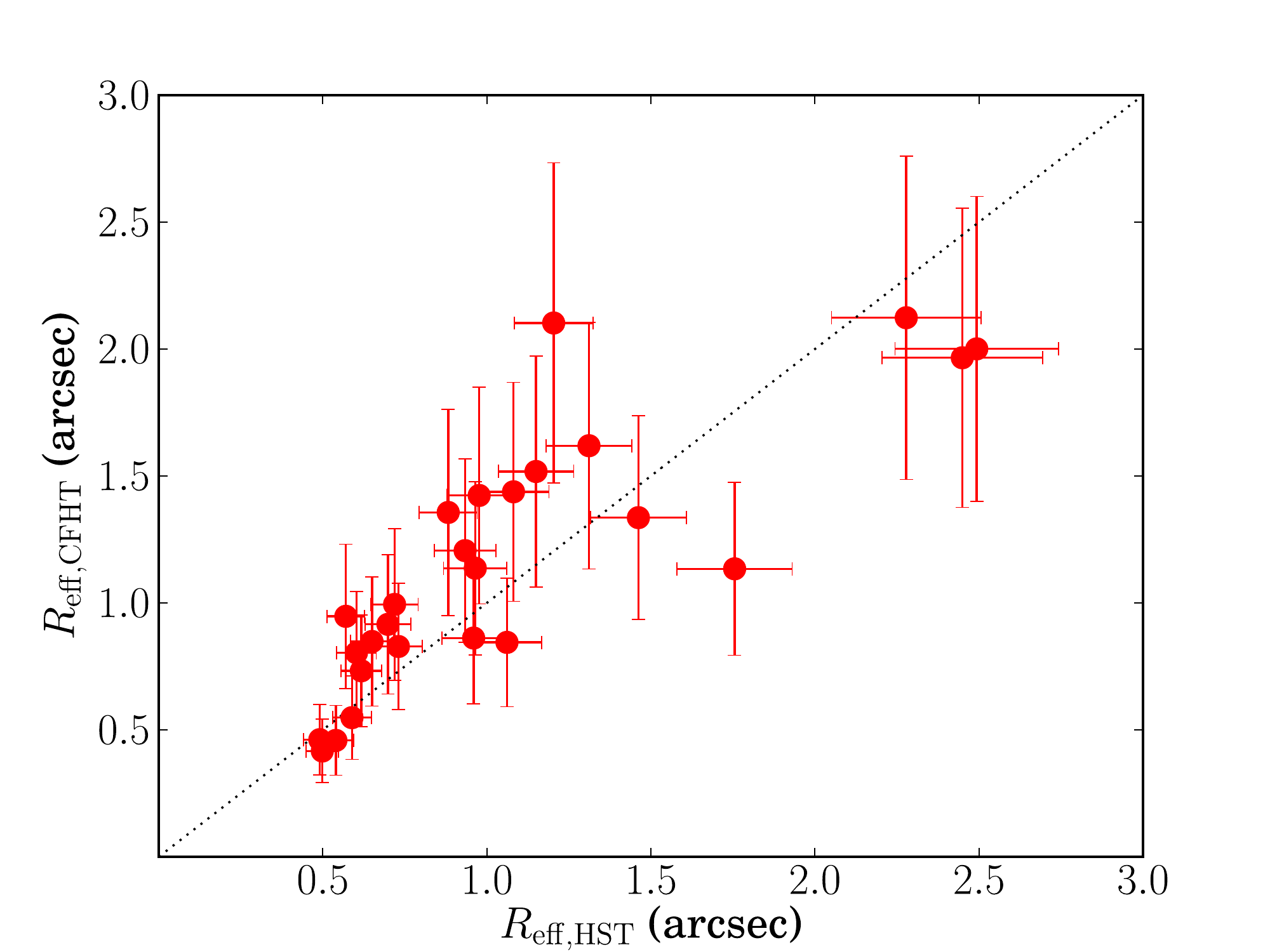}
\caption{\label{fig:reff}Comparison between effective radii measured from ground-based versus space-based photometry. Error bars on \hst effective radii represent the assigned 10\% systematic uncertainty due to fixing the light profile to a de Vaucouleurs model. The relative scatter between the best fit values of the two measurements is 30\% and is used to quantify uncertainties in CFHT effective radii.
}
\end{figure}

Uncertainties on the \hst\ effective radii are dominated by the choice of the model light profile: different models can fit the data equally well but give different estimates of $R_{\mathrm{eff}}$.
The dispersion is $\sim10\%$, estimated by repeating the fit with a different surface brightness model, a Hernquist profile, and comparing the newly obtained values of $R_{\mathrm{eff}}$ with the de Vaucouleurs ones.
Uncertainties on the CFHT effective radii are instead dominated by contamination from the background sources.
Effective radii measured from CFHT images are in good agreement
with those measured from \hst\ data, when present, as shown
in \Fref{fig:reff}. The scatter on the quantity
$R_{\mathrm{eff,CFHT}} - R_{\mathrm{eff,HST}}$ is $\sim30\%$; we take this as our uncertainty on CFHT effective radii.

% %%%%%%%%%%%%%%%%%%%%%%%%%%%%%%%%%%%%%
\renewcommand{\arraystretch}{1.10} 
\begin{deluxetable*}{lccccccccccc}
\tablewidth{0pt}
\tabletypesize{\small}
\tablecaption{Lens light parameters, CFHT photometry.\label{table:cfhtphot}}
\tabletypesize{\footnotesize}
\tablehead{
\colhead{Name} & \colhead{$R_{\mathrm{eff}}$} & \colhead{$q$} & \colhead{PA} &
\colhead{$u$} & \colhead{$g$} & \colhead{$r$} & \colhead{$i$} & \colhead{$z$} & \colhead{$J$} & \colhead{$H$} & \colhead{$Ks$} \\
& (arcsec) & & (degrees) & & & & & & & &
}
\startdata
SL2SJ020833-071414 & $1.06$ & $0.81$ & $61.1$ & $22.71$ & $20.64$ & $18.99$ & $18.22$ & $17.90$ & & &\\ 
SL2SJ021206-075528 & $0.78$ & $0.79$ & $-29.2$ & $23.33$ & $21.32$ & $19.75$ & $18.90$ & $18.61$ & & &\\ 
SL2SJ021247-055552 & $1.22$ & $1.00$ & $-9.1$ & $23.59$ & $22.73$ & $21.44$ & $20.21$ & $19.77$ & & &\\ 
SL2SJ021325-074355 & $1.97$ & $0.60$ & $21.2$ & $24.29$ & $22.28$ & $20.78$ & $19.27$ & $18.82$ & & & $17.43$\\ 
SL2SJ021411-040502 & $1.21$ & $0.88$ & $57.1$ & $23.82$ & $22.39$ & $20.88$ & $19.65$ & $19.23$ & $18.55$ & & $17.87$\\ 
SL2SJ021737-051329 & $0.73$ & $0.90$ & $87.6$ & $23.21$ & $22.17$ & $20.92$ & $19.70$ & $19.33$ & $18.72$ & & $17.97$\\ 
SL2SJ021801-080247 & $1.02$ & $1.00$ & $-49.8$ & $23.05$ & $22.07$ & $21.32$ & $20.33$ & $19.64$ & & &\\ 
SL2SJ021902-082934 & $0.95$ & $0.74$ & $82.6$ & $23.02$ & $21.37$ & $19.70$ & $18.94$ & $18.55$ & & & $17.59$\\ 
SL2SJ022046-094927 & $0.53$ & $0.71$ & $-68.5$ & $24.17$ & $22.33$ & $20.88$ & $19.88$ & $19.52$ & & &\\ 
SL2SJ022056-063934 & $1.42$ & $0.54$ & $-74.8$ & $21.65$ & $19.85$ & $18.47$ & $17.86$ & $17.59$ & & &\\ 
SL2SJ022346-053418 & $1.31$ & $0.59$ & $63.4$ & $22.93$ & $21.09$ & $19.56$ & $18.70$ & $18.29$ & & &\\ 
SL2SJ022357-065142 & $1.36$ & $0.95$ & $37.2$ & $23.13$ & $21.03$ & $19.42$ & $18.63$ & $18.30$ & & & $17.45$\\ 
SL2SJ022511-045433 & $2.12$ & $0.72$ & $27.5$ & $20.32$ & $18.58$ & $17.36$ & $16.81$ & $16.55$ & $15.99$ & $15.64$ & $15.48$\\ 
SL2SJ022610-042011 & $0.84$ & $0.87$ & $52.0$ & $23.30$ & $21.28$ & $19.70$ & $18.80$ & $18.46$ & $18.09$ & $17.70$ & $17.38$\\ 
SL2SJ022648-040610 & $0.48$ & $0.30$ & $-47.5$ & $25.12$ & $23.26$ & $21.57$ & $20.12$ & $19.57$ & $18.90$ & $18.52$ & $18.10$\\ 
SL2SJ022648-090421 & $1.40$ & $0.81$ & $56.8$ & $22.65$ & $20.46$ & $18.79$ & $18.06$ & $17.69$ & & &\\ 
SL2SJ023251-040823 & $1.14$ & $0.70$ & $-72.6$ & $22.28$ & $20.71$ & $19.31$ & $18.72$ & $18.44$ & & & $17.30$\\ 
SL2SJ084847-035103 & $0.45$ & $0.82$ & $-65.4$ & $24.57$ & $23.57$ & $22.16$ & $20.81$ & $20.39$ & & &\\ 
SL2SJ084909-041226 & $0.46$ & $0.51$ & $43.7$ & $24.90$ & $23.16$ & $21.70$ & $20.16$ & $19.70$ & & & $18.60$\\ 
SL2SJ084934-043352 & $1.59$ & $0.78$ & $36.4$ & $22.52$ & $20.49$ & $19.01$ & $18.31$ & $18.02$ & & &\\ 
SL2SJ084959-025142 & $1.34$ & $0.79$ & $-65.4$ & $21.75$ & $19.85$ & $18.56$ & $17.94$ & $17.68$ & & & $16.63$\\ 
SL2SJ085019-034710 & $0.28$ & $0.22$ & $ 1.2$ & $23.52$ & $21.48$ & $20.07$ & $19.38$ & $19.14$ & & &\\ 
SL2SJ085327-023745 & $1.47$ & $0.81$ & $-24.3$ & $23.07$ & $22.24$ & $21.46$ & $20.29$ & $19.78$ & & &\\ 
SL2SJ085540-014730 & $0.69$ & $0.82$ & $-70.8$ & $22.80$ & $21.42$ & $20.05$ & $19.37$ & $19.10$ & & &\\ 
SL2SJ085559-040917 & $1.13$ & $0.82$ & $23.1$ & $23.18$ & $21.10$ & $19.48$ & $18.72$ & $18.35$ & & &\\ 
SL2SJ085826-014300 & $0.55$ & $0.77$ & $-86.2$ & $24.09$ & $23.15$ & $21.85$ & $20.78$ & $20.38$ & & & $19.20$\\ 
SL2SJ090106-025906 & $0.42$ & $0.82$ & $-67.5$ & $24.53$ & $23.81$ & $22.40$ & $21.16$ & $20.73$ & & & $19.80$\\ 
SL2SJ090407-005952 & $2.00$ & $0.64$ & $71.1$ & $23.59$ & $21.61$ & $20.57$ & $19.47$ & $19.12$ & & & $17.71$\\ 
SL2SJ095921+020638 & $0.46$ & $0.90$ & $42.0$ & $25.28$ & $22.74$ & $21.23$ & $20.23$ & $19.90$ & $19.72$ & $19.38$ & $19.13$\\ 
SL2SJ135847+545913 & $0.92$ & $0.79$ & $-72.4$ & $23.93$ & $21.66$ & $20.14$ & $19.16$ & $18.78$ & & &\\ 
SL2SJ135949+553550 & $1.13$ & $0.67$ & $35.7$ & $24.40$ & $23.30$ & $21.90$ & $20.69$ & $20.04$ & & &\\ 
SL2SJ140123+555705 & $0.86$ & $0.75$ & $-41.9$ & $23.84$ & $21.64$ & $20.05$ & $18.97$ & $18.57$ & & &\\ 
SL2SJ140156+554446 & $1.44$ & $0.82$ & $20.2$ & $23.07$ & $20.83$ & $19.28$ & $18.47$ & $18.02$ & & &\\ 
SL2SJ140221+550534 & $1.52$ & $0.94$ & $-18.1$ & $22.28$ & $20.47$ & $18.89$ & $18.19$ & $17.82$ & & &\\ 
SL2SJ140454+520024 & $2.03$ & $0.79$ & $67.2$ & $22.37$ & $20.17$ & $18.56$ & $17.73$ & $17.37$ & & &\\ 
SL2SJ140533+550231 & $0.56$ & $0.67$ & $15.8$ & 24.32 & 22.48 & 21.10 & 20.13 & 19.62 & & & \\ 
 & $1.11$ & $0.98$ & $-36.03$ & 23.14 & 22.28 & 21.11 & 20.23 & 19.73 & & & \\ 
SL2SJ140546+524311 & $0.83$ & $0.89$ & $-27.5$ & $23.73$ & $21.62$ & $20.10$ & $19.10$ & $18.74$ & & &\\ 
SL2SJ140614+520253 & $2.21$ & $0.50$ & $-60.9$ & $22.67$ & $20.64$ & $19.08$ & $18.22$ & $17.85$ & & &\\ 
SL2SJ140650+522619 & $0.80$ & $0.67$ & $87.4$ & $23.84$ & $22.59$ & $21.31$ & $19.96$ & $19.47$ & & &\\ 
SL2SJ141137+565119 & $0.85$ & $0.99$ & $ 2.5$ & $21.68$ & $20.37$ & $19.08$ & $18.55$ & $18.27$ & & &\\ 
SL2SJ141917+511729 & $2.10$ & $0.64$ & $47.7$ & $23.33$ & $20.96$ & $19.43$ & $18.56$ & $18.21$ & & &\\ 
SL2SJ142003+523137 & $0.72$ & $0.20$ & $71.2$ & $24.69$ & $22.85$ & $21.43$ & $20.70$ & $20.31$ & & &\\ 
SL2SJ142031+525822 & $1.11$ & $0.62$ & $-86.0$ & $22.94$ & $20.88$ & $19.36$ & $18.68$ & $18.33$ & & &\\ 
SL2SJ142059+563007 & $1.62$ & $0.54$ & $-12.6$ & $22.72$ & $20.73$ & $19.29$ & $18.51$ & $18.16$ & & &\\ 
SL2SJ142321+572243 & $1.42$ & $0.82$ & $62.9$ & $23.67$ & $21.65$ & $20.00$ & $18.95$ & $18.64$ & & &\\ 
SL2SJ142731+551645 & $0.39$ & $0.31$ & $-63.7$ & $23.33$ & $22.00$ & $20.74$ & $19.85$ & $19.47$ & & &\\ 
SL2SJ220329+020518 & $0.99$ & $0.81$ & $-44.7$ & $22.68$ & $21.08$ & $19.80$ & $19.15$ & $18.83$ & & & $17.85$\\ 
SL2SJ220506+014703 & $0.66$ & $0.48$ & $87.2$ & $23.71$ & $21.69$ & $20.09$ & $19.24$ & $18.92$ & & & $17.73$\\ 
SL2SJ220629+005728 & $1.37$ & $0.63$ & $-25.1$ & $23.59$ & $22.31$ & $20.99$ & $19.75$ & $19.24$ & & & $17.66$\\ 
SL2SJ221326-000946 & $0.27$ & $0.34$ & $-29.1$ & $23.60$ & $21.78$ & $20.33$ & $19.74$ & $19.44$ & & & $18.61$\\ 
SL2SJ221407-180712 & $0.68$ & $0.72$ & $57.0$ & $24.81$ & $22.81$ & $21.37$ & $20.15$ & $19.73$ & & &\\ 
SL2SJ221852+014038 & $0.90$ & $0.53$ & $-67.1$ & $23.86$ & $21.70$ & $20.19$ & $19.07$ & $18.67$ & & & $17.54$\\ 
SL2SJ221929-001743 & $1.01$ & $0.78$ & $85.1$ & $21.32$ & $19.52$ & $18.31$ & $17.78$ & $17.50$ & & &\\ 
SL2SJ222012+010606 & $0.80$ & $0.87$ & $-22.4$ & $22.34$ & $20.47$ & $19.38$ & $18.84$ & $18.56$ & & & $17.69$\\ 
SL2SJ222148+011542 & $1.12$ & $0.81$ & $79.6$ & $22.02$ & $20.15$ & $18.83$ & $18.21$ & $17.91$ & & & $16.90$\\ 
SL2SJ222217+001202 & $1.56$ & $0.66$ & $37.7$ & $22.77$ & $21.03$ & $19.62$ & $18.88$ & $18.53$ & & &\\ 

\enddata
\tablecomments{Best fit parameters for de Vaucouleurs models of the surface brightness profile of the lens galaxies, after careful manual masking of the lensed images.  Columns 2--4 correspond to the effective radius ($R_{\rm eff}$), the axis ratio of the elliptical isophotes ($q$), and the position angle measured east of north (PA).  
The system SL2SJ140533+550231 has two lens galaxies of comparable magnitude, and the parameters of both galaxies are given. 
The typical uncertainties are a few degrees for the position angle, $\Delta q\sim0.03$ for the axis ratio, $0.3$ for $u$-band magnitudes, $0.2$ for $g$ and $r$-band magnitudes, $0.1$ for magnitudes in the remaining bands, $30\%$ on the effective radii.
}
\end{deluxetable*}
% %%%%%%%%%%%%%%%%%%%%%%%%%%%%%%%%%%%%%

% %%%%%%%%%%%%%%%%%%%%%%%%%%%%%%%%%%%%%
\begin{deluxetable*}{ccccccccc}
\tablewidth{0pt}
\tablecaption{Lens light parameters, \hst\ photometry.\label{table:hstphot}}
\tablehead{
\colhead{Name} & \colhead{$R_{\mathrm{eff}}$} & \colhead{$q$} & \colhead{PA} &
\colhead{$m_{\mathrm{F814W}}$} & \colhead{$m_{\mathrm{F606W}}$} & \colhead{$m_{\mathrm{F600LP}}$} & \colhead{$m_{\mathrm{F475X}}$} & Set \\
& (arcsec) & & (degrees) & & & & &
}
\startdata
SL2SJ020833-071414 & $0.94$ & $0.79$ & $70.5$ & $\cdots$ & $\cdots$ & $18.58$ & $20.69$ & (c) \\ 
SL2SJ021325-074355 & $2.45$ & $0.67$ & $34.9$ & $\cdots$ & $21.08$ & $\cdots$ & $\cdots$ & (b) \\ 
SL2SJ021411-040502 & $0.93$ & $0.91$ & $77.8$ & $19.07$ & $20.73$ & $\cdots$ & $\cdots$ & (a) \\ 
SL2SJ021737-051329 & $0.61$ & $0.91$ & $71.7$ & $19.13$ & $20.77$ & $\cdots$ & $\cdots$ & (a) \\ 
SL2SJ021902-082934 & $0.57$ & $0.73$ & $73.2$ & $\cdots$ & $\cdots$ & $19.55$ & $21.79$ & (c) \\ 
SL2SJ022357-065142 & $0.88$ & $0.81$ & $48.0$ & $\cdots$ & $\cdots$ & $19.21$ & $21.43$ & (c) \\ 
SL2SJ022511-045433 & $2.28$ & $0.72$ & $25.1$ & $\cdots$ & $17.68$ & $\cdots$ & $\cdots$ & (b) \\ 
SL2SJ022610-042011 & $1.06$ & $0.81$ & $61.9$ & $\cdots$ & $20.00$ & $\cdots$ & $\cdots$ & (b) \\ 
SL2SJ022648-040610 & $1.10$ & $0.38$ & $-47.4$ & $\cdots$ & $21.45$ & $\cdots$ & $\cdots$ & (b) \\ 
SL2SJ023251-040823 & $0.96$ & $0.74$ & $-68.2$ & $\cdots$ & $19.91$ & $\cdots$ & $\cdots$ & (b) \\ 
SL2SJ084909-041226 & $0.49$ & $0.49$ & $39.1$ & $\cdots$ & $\cdots$ & $20.62$ & $23.40$ & (c) \\ 
SL2SJ084959-025142 & $1.46$ & $0.78$ & $-64.1$ & $\cdots$ & $18.80$ & $\cdots$ & $\cdots$ & (b) \\ 
SL2SJ085826-014300 & $0.59$ & $0.81$ & $82.2$ & $\cdots$ & $22.22$ & $\cdots$ & $\cdots$ & (b) \\ 
SL2SJ090106-025906 & $0.50$ & $0.80$ & $-20.3$ & $\cdots$ & $22.78$ & $\cdots$ & $\cdots$ & (b) \\ 
SL2SJ090407-005952 & $2.50$ & $0.79$ & $74.4$ & $\cdots$ & $20.89$ & $\cdots$ & $\cdots$ & (b) \\ 
SL2SJ095921+020638 & $0.54$ & $0.78$ & $26.0$ & $\cdots$ & $21.65$ & $\cdots$ & $\cdots$ & (b) \\ 
SL2SJ135847+545913 & $0.70$ & $0.80$ & $-57.1$ & $\cdots$ & $\cdots$ & $19.65$ & $21.99$ & (c) \\ 
SL2SJ135949+553550 & $1.76$ & $0.61$ & $30.7$ & $\cdots$ & $22.01$ & $\cdots$ & $\cdots$ & (b) \\ 
SL2SJ140123+555705 & $0.96$ & $0.78$ & $-43.5$ & $\cdots$ & $20.33$ & $\cdots$ & $21.66$ & (b,d) \\ 
SL2SJ140156+554446 & $1.08$ & $0.79$ & $27.2$ & $\cdots$ & $\cdots$ & $18.90$ & $21.10$ & (c) \\ 
SL2SJ140221+550534 & $1.15$ & $0.87$ & $-43.9$ & $\cdots$ & $\cdots$ & $18.59$ & $20.63$ & (c) \\ 
SL2SJ140533+550231 & $0.91$ & $0.75$ & $-21.3$ & $\cdots$ & 21.75 & $\cdots$ & $\cdots$ & (b) \\ 
 & $0.41$ & $0.93$ & $-12.72$ & $\cdots$ & 22.02 & $\cdots$ & $\cdots$ &  \\ 
SL2SJ140546+524311 & $0.73$ & $0.82$ & $-27.8$ & $\cdots$ & $\cdots$ & $19.44$ & $21.77$ & (c) \\ 
SL2SJ140650+522619 & $0.60$ & $0.65$ & $84.3$ & $\cdots$ & $21.97$ & $\cdots$ & $\cdots$ & (b) \\ 
SL2SJ141137+565119 & $0.65$ & $0.79$ & $ 8.1$ & $\cdots$ & $\cdots$ & $18.93$ & $20.65$ & (c) \\ 
SL2SJ141917+511729 & $1.20$ & $0.69$ & $48.8$ & $\cdots$ & $\cdots$ & $19.24$ & $21.54$ & (c) \\ 
SL2SJ142031+525822 & $1.04$ & $0.65$ & $-80.0$ & $\cdots$ & $19.86$ & $\cdots$ & $\cdots$ & (b) \\ 
SL2SJ142059+563007 & $1.31$ & $0.58$ & $-13.6$ & $\cdots$ & $\cdots$ & $18.96$ & $21.01$ & (c) \\ 
SL2SJ142321+572243 & $0.98$ & $0.85$ & $68.2$ & $\cdots$ & $\cdots$ & $19.48$ & $22.01$ & (c) \\ 
SL2SJ142731+551645 & $0.50$ & $0.29$ & $-62.1$ & $\cdots$ & $21.08$ & $\cdots$ & $\cdots$ & (b) \\ 
SL2SJ220329+020518 & $0.72$ & $0.84$ & $-50.6$ & $\cdots$ & $\cdots$ & $19.79$ & $21.76$ & (c) \\ 
SL2SJ221326-000946 & $0.50$ & $0.38$ & $-32.5$ & $\cdots$ & $20.64$ & $\cdots$ & $\cdots$ & (b) \\ 
SL2SJ221407-180712 & $0.77$ & $0.69$ & $49.1$ & $\cdots$ & $21.69$ & $\cdots$ & $\cdots$ & (b) \\ 

\enddata
\tablecomments{Best fit parameters for de Vaucouleurs models of the surface brightness profile of the lens galaxies, after careful manual masking of the lensed images.  Columns 2--4 correspond to the effective radius ($R_{\rm eff}$), the axis ratio of the elliptical isophotes ($q$), and the position angle measured east of north (PA).  
The system SL2SJ140533+550231 has two lens galaxies of comparable magnitude, and the parameters of both galaxies are given. The last column indicates the set of observations used, from the list in \Tref{table:HST}.
}
\end{deluxetable*}
% %%%%%%%%%%%%%%%%%%%%%%%%%%%%%%%%%%%%%

% %%%%%%%%%%%%%%%%%%%%%%%%%%%%%%%%%%%%%
% %%%%%%%%%%%%%%%%%%%%%%%%%%%%%%%%%%%%%

% - - - - - - - - - - - - - - - - - - - - - - - - - - - - - - - - - - - - - - - 

\section{Lens models}\label{sect:lensing}

The main goal is to measure Einstein radii of our lenses. 
We define the Einstein radius $R_{\rm{Ein}}$ to be the radius within which the
mean surface mass density $\bar{\Sigma}(<R_{\rm{Ein}})$ equals the critical density
$\Sigma_{\rm{cr}}$ of the lensing configuration. While the critical
density depends on the lens and source redshifts, the ratio of $\bar{\Sigma}(<R_{\rm{Ein}})/\Sigma_{\rm{cr}}$ (i.e., the convergence) does not: in practice then, the
deflection angles and lensed image positions can all be predicted given a model with its Einstein radius in angular units. We only consider Einstein radii in
angular units throughout this paper.

\subsection{The method}
\label{sec:REin}

We measure Einstein radii by fitting model mass distributions to the lensing data.
We describe our lenses as singular isothermal ellipsoids (SIE), with convergence~$\kappa$ given by
\begin{equation}
\kappa(x,y) = \frac{R_{\rm{Ein}}}{2r},
\end{equation}
where $r^2 \equiv qx^2 + y^2/q$ and $q$ is the axis ratio of the elliptical isodensity contours.
The free parameters of the lens model are therefore $R_{\rm{Ein}}$, the axis ratio $q$, the position angle (PA) of the major axis, and the $x$ and $y$ positions of the centroid.
In principle, more degrees of freedom could be introduced. 
In some cases, lens models are found to require a constant external shear, with strength $\gamma_{ext}$ and position angle ${\rm PA}_{\rm ext}$, in order  to describe the lensing effect of massive objects
(such as groups or clusters) close to the optical axis. However, this external
shear is highly degenerate with the mass orientation of the main lens, and our
data are not detailed enough to distinguish between the two. For this reason we
only include a shear component for the lenses that we cannot otherwise find a working model.

The fit is performed by generating model lensed images and comparing them to the observed images that have the lens light subtracted.
For fixed lens parameters, light from the image plane is mapped back to a grid on the source plane and the source light distribution is then reconstructed following \citet{Suy++06}.
This source reconstruction, as well as the entire lensing analysis, follows a Bayesian approach.
For a given model lens, the Bayesian evidence of the source reconstruction is computed, which then defines the %likelihood
quality of the lens model.
The lens parameter space is then explored with a Monte-Carlo Markov Chain (MCMC) sampler, propagating the source reconstruction evidence as the likelihood of the lens model parameters.

The practical realization of this procedure is done by using the lens modeling software {\tt GLEE}, developed by \citet{S+H10}.
This approach differs slightly from the one adopted in Paper I, in that a pixelized source reconstruction is used instead of fitting S\'{e}rsic components.
To make sure that our analysis is robust, we repeat the fit for the systems previously analyzed in Paper I.
This allows us to gauge the importance of systematic effects related to the choice of modeling technique.

For systems with \hst\ imaging in more than one band, only the bluest band image is used for the analysis as the signal from the blue star-forming lensed sources is highest. The $g$ band image is used when modeling CFHT data.
Typically we only attempt to model a small region of the image around the identified lensed sources, then check that our lens models do not predict detectable lensed sources in areas outside the data region.
We assume uniform priors on all the lens parameters except the centroid, for which we use a Gaussian PDF centered on the observed light distribution and with a dispersion of 1 pixel.
For systems with only ground-based imaging, for which the lensing signal is diluted by the large PSF, we keep the centroid fixed to that of the optimal light profile. In some cases we also adopt a Gaussian prior on the mass PA, centered on the PA of the light, or we keep the PA fixed.
Those cases are individually discussed below.

Our analysis also allows us to determine the brightness of the lensed sources.
This is important information as it allows us to constrain their distance in cases where their spectroscopic redshift is unknown \citep{Ruf++11}.
The unlensed magnitude of the background object is recovered by fitting S\'{e}rsic components to the reconstructed source.

The values of the measured lens parameters with 68\% credible intervals ($1-\sigma$ uncertainties) derived from the posterior probability distribution function marginalized over the remaining parameters are reported in \Tref{table:lensfit}. 
Cutouts of the lens systems with the most-probable image and source reconstruction are shown in \Fref{fig:lenscuts}.
All images are orientated north up and east left, with the exception of lens models based on WFPC2 data. Those models are performed in the native detector frame in order to avoid degrading further the quality of the WFPC2 images, as they typically have a low S/N.
In such situations a compass is displayed to guide the eye.

The formal uncertainties on the Einstein radius given by the MCMC sampling are typically very small: the 1-$\sigma$ uncertainty is for most lenses smaller than $1\%$.
However, our measurements of the Einstein radius rely partly on the assumption of an SIE profile for the lens mass distribution:
in principle, mass models with density slope different from isothermal or isodensity contours different from ellipses can produce different Einstein radii.
Perhaps more significantly, some systematic effects can be introduced at various points in our analysis: in particular, the assertions of a specific model PSF, a specific arc mask, and a specific lens light subtraction procedure all induce uncertainty in the final prepared data image \citep[\eg][]{Mar++07, Suy++09}. 
\citet{Bol++08a} estimated the systematic uncertainty on typical Einstein radius measurements to be about 2\%.
We can further verify this result by comparing Einstein radius measurements from paper I with the new values found here.
The analysis of Paper I differs from the present one in the lens light subtraction, choice of the arc mask and lens model technique (S\'{e}rsic component fitting versus pixelized source reconstruction), so a comparison of the two different measurements should reflect systematics from most of the effects listed above.
For a few of the systems already analyzed in Paper I, the current lens models are qualitatively different from the ones presented in Paper I and the measured values of the Einstein radii are correspondingly different.
In most cases this is in virtue of the collection of new data with \hst\ WFC3 that revealed features on the lensed arcs, previously undetected, that helped improve the lens model.
After excluding those systems, the relative scatter between the most probable values of $R_{\mathrm{Ein}}$ measured in the two different approaches (current and previous) is 3\%.
We thus take $3\%$ as our estimate of the systematic uncertainty on the measurement of the Einstein radius with the technique used here, and convolve the posterior probability distribution for the Einstein radius obtained from the MCMC with a Gaussian with $3\%$ dispersion.
All the uncertainties on $R_{\rm{Ein}}$ quoted in this paper reflect this choice.

\subsection{The lenses}\label{ssect:lenses}

Although the lens modeling procedure is the same for all lenses, each system has its own peculiarities that need to be taken care of.
In what follows we describe briefly and case by case the relevant aspects of those lens models that deserve some discussion.
%While many of our targets show at a glance a clear lens-like morphology, it is in some cases difficult to establish with absolute certainty whether a system is a gravitational lens or not, even when we are able to model it as such.
%It is nevertheless very important to quantify the likelihood for an observed system of being a strong lens.
%We do it with the following scheme, standard in the strong lensing community.

Lens grades are also discussed in this subsection, when explanation is needed, and are reported in \Tref{table:lensfit}.
In general we apply the following guidelines. For a system with \hst\ imaging we require, in order for it to be a grade A, that at least a pair of multiple images of the same source is visible and that we can describe it with a robust lens mass model compatible with the light distribution of the lens galaxy (i.e. similar centroid, orientation and axis ratio).
For systems with only ground-based imaging we impose the additional requirement of having a spectroscopic detection of the background source, in order to be sure that the blue arcs that we observe are not part of the foreground galaxy.
Spectroscopic data therefore enters the lens classification process. We refer to our companion paper (Paper IV) when discussing spectroscopic measurements.
%\QUERY{PJM}{What determines grade B over grade C? And grade C over X?}
Furthermore, systems with a reliable ground-based lens model but no source spectroscopy are given grade B, as well as systems with secure spectroscopic detection of the source but no robust lens model.
Systems lacking both, or for which we suspect that strong lensing might not be present are instead given grade C.
We stress that a grade is not necessarily a statement on the quality or usefulness of a system as a lens, but rather its likelihood of being a strong lens given the available data.
Consequently, grades are subject to change as new data become available.

\begin{itemize}

\item SL2SJ020833-071414. The lensed features of this system consist of a double image of a bright compact component and a low surface brightness ring.
The model cannot fully reproduce the image of the bright double but this is probably due to the presence of a compact unresolved component, like an AGN. AGNs in the source plane are difficult to model with a pixelized reconstruction technique, because the image regularization process smoothes our model images. This effect is present in other lenses with sharp peaks in the source surface brightness distribution. 
Since the signal-to-noise ratio of the \hst\ image is low and no additional information comes from spectroscopy, this lens is given a grade B.

\item SL2SJ021325-074355.
The source lensed by this high redshift galaxy ($z_d = 0.717$) appears to have two separate bright components.
Our source reconstruction confirms this picture.
There is a massive elliptical galaxy in the foreground that may be providing extra deflection to the light coming from the source, thus perturbing the image.
This perturber is very close to the observer ($z=0.0161$, from SDSS) and therefore its lensing power is greatly reduced with respect to the main deflector.
We model the mass distribution from this galaxy with an additional SIE with centroid and PA fixed to the light distribution and $R_{Ein}$ and $q$ as free parameters.
To quickly calculate the deflection angles from this perturber we make the approximation that it lies at the same redshift as the main lens.
While this is not formally correct, the model still describes qualitatively the presence of an extra source of deflection towards the direction of the foreground galaxy.
The impact of this perturber on the lensing model is in any case
small. 

\item SL2SJ021411-040502.
The source has two bright components, one of which is lensed into the big arc and its fainter couter-image. The second component forms a double of smaller magnification.
This lens was modeled in Paper I where we explained how there are two lens models, with different Einstein radii, that match the image configuration. 
The pixelized source reconstruction technique adopted here to model the lens favors the solution alternative to the one chosen in Paper I.

\item SL2SJ021737-051329.
This lens system is in a cusp configuration, meaning that the source lies just within one of the four cusps of the astroid caustic of the lens.
Either a mass centroid offset from the light center or a large shear
is required to match the curvature of the big arc. This was also
needed in Paper I and previously found by \citet{Tu++09}. Here, we
find the amount of external shear to be $\gamma_{\mathrm{ext}} = 0.11\pm0.01$ 

\item SL2SJ021801-010606. This system shows a nearly complete ring. The redshift of the blue component is 2.06 but we were not able to measure the redshift of the deflector, therefore we label this system as a grade B lens, needing follow-up with deeper spectroscopy.

\item SL2SJ022346-053418. The CFHT image shows an extended arc and a bright knot at the opposite side with respect to the lens.
Although this latter component might be the counter-image to the arc, its color is different and it is not detected spectroscopically.
Therefore only the arc is used for the lensing analysis.
The main arc has a higher redshift than the lens, however the lens model is not definitive in assessing whether this system is a strong lens. This is therefore a grade B lens.

\item SL2SJ022357-065142. The lensed source appears to have a complex
morphology. We identify three distinct components, each of which is doubly
imaged.

%\item SL2SJ084847-035103. The blue arc is at a different redshift than
%the lens. However the system
%lacks a clear counterimage (at least in the CFHT data). It is therefore a grade B.

\item SL2SJ084934-043352. 
Only one arc is visible in the CFHT image.
In order to obtain a meaningful lens model we need to fix the PA of the mass profile to that of the light. This system is a grade B due to the lack of spectroscopic detection of the background source.

\item SL2SJ084959-025142 is a double-like lens system. Part of the light close
to the smaller arc is masked out in our analysis, as it is probably a
contamination from objects not associated with the lensed
source. 

\item SL2SJ085019-034710. The CFHT image shows a bright arc produced by the
lensing effect of a disk galaxy. The lens model predicts the presence of a
counter-image opposite to the arc, but it is not bright enough to be
distinguished from the disk of the lens. In addition, such a
counterimage could suffer from substantial extinction. 

%\item SL2SJ085327-023745. The two main blue components of this system are spectroscopically detected to be at the same redshift and at redshift higher than the lens. However the lens model leaves non-negligible residuals in the image and is therefore not a robust measurement, making this a grade B lens.

\item SL2SJ085559-040917.
The main blue arc of this system is at redshift 2.95. 
The other blue features seen in CFHT data however are too faint for us to establish an unambiguous interpretation of the lens configuration. Therefore we conservatively assign grade B to this system. Higher resolution photometry is needed to confirm this lens.

%\item SL2SJ090106-025906. Only a small arc and a faint blob are
%  visible in the low S/N WFPC2 image of this system. The arc is spectroscopically measured to be at a higher redshift than the lens while the blob is not detected. Assuming that
%  they are multiple images we do not get a working model, as the best
%  such model predicts other images that are not observed. If we only
%  model the main arc we cannot put meaningful constraints on the lens
%  model parameters. Grade B. 

\item SL2SJ090106-025906.
The WFPC2 image of this system is contaminated with a cosmic ray, which has been masked out in our analysis. Our lens model predicts an image at the position of the cosmic ray, the presence of which cannot be verified with our data. The model however appears to be convincing and the background source is spectroscopically detected, therefore this is a grade A lens.

\item SL2SJ095921+020638. This system, belonging to the COSMOS survey
  had previously been modeled by \citet{Ang++09}. These authors report
  a source redshift of $3.14\pm0.05$ whereas we find a slightly
  greater value of $3.35\pm0.01$ based on our own XSHOOTER data (Paper
  IV). They report an Einstein radius $\REin\sim0\farcs71$ in close
  agreement with our $0\farcs74 \pm 0\farcs04$ estimate. 

\item SL2SJ135847+545913. We identify two distinct bright components in the
source: one forms the big arc, the other one is only doubly-imaged.

%\item SL2SJ135949+553550 is a cusp-like system with a quadruply-imaged bright
%compact component.

\item SL2SJ140123+555705 is a cusp-like system: three images of a single
bright knot can be identified on the arc. The counter-image however is too
faint to be detected in the WFC3 snapshot.
This lens was already modeled in Paper I. The Einstein radius that we
obtain here is not consistent with the value reported then. This is
because the current model is obtained by analyzing newly obtained WPC3
data, which reveal more details on the arc. The lack of a counterimage
does not prevent an accurate lens modeling because the main arc is
very thin, curved and extended. 

\item SL2SJ140533+550231. This is a particular system, in that there are two
lens galaxies of comparable brightness. The lensed image shows four images of
a bright knot. We model the system with two SIE components, centered in
correspondence with the two light components. Our inference shows a
substantial degeneracy between the Einstein radii of the two
lenses. 

\item SL2SJ140546+524311. This system shows a quadruply-imaged bright compact
component. Two of the images are almost merged. A relatively large shear is
required to match the position and shape of the counter-image opposite to the
arcs.

\item SL2SJ140614+520253. A few different blue blobs are visible in
  the CFHT image, but there is no working lens model that can
  associate them with the same source. As done in Paper I, we model
  only the bright extended arc. The resulting Einstein radius differs
  from the value of Paper I. This reflects the fact that the
  interpretation of this system as a lens is not
  straightforward.  This is Grade B until future \hst\ data shed more light on the
  actual nature of this system.

\item SL2SJ140650+522619 has a cusp configuration. Even though the source
appears to have two separate components, the compact structure outside of the
main arc is actually a foreground object, as revealed by our spectroscopic
observations.

\item SL2SJ141137+565119 shows a complete ring. Our lens model cannot account
for all the flux in one bright knot on the arc, North of the lens. 
This could
be the result of the presence of substructure close to the highly
magnified unresolved knot that requires a minute knowledge of the PSF.

\item SL2SJ141917+511729. Only two bright points can be identified on the arc
of this system, while no counter-image is visible. The Einstein radius of this
lens is rather large ($\sim 4\arcsec$), which puts this system in the category of
group-scale lenses.

\item SL2SJ142003+523137. This disk galaxy is producing a lensed arc. The
reconstructed source is compact and difficult to resolve. The predicted
counter-image of the arc is too faint to be detected and possibly
affected by extinction. 

\item SL2SJ142059+563007. The WFC3 image of this lens offers a detailed view on
the source structure. We identify three separate bright components, two quads
and one double, which allow us to constrain robustly the Einstein radius.

\item SL2SJ142731+551645. The source lensed by this disky galaxy is in a
typical fold-like configuration, with two of its four images merging into an
arc.

\item SL2SJ220329+020518. This system shows a bright arc, and a possible
counter-image close to the center. However, we are not able to fit both the
light from the arc and the candidate counter-image. On the other hand, our
spectroscopic analysis reveals OII emission at the redshift of the lens (Paper
IV), which suggests that the blue bright spot close to the center might be a
substructure associated with the lens and not the source. We model the system
using light from the arc only. Our model predicts the existence of a faint
counter-image that cannot be ruled out by our snapshot
observation. 

\item SL2SJ220506+014703. The spectroscopic follow-up revealed emission from the bright arc at $z=2.52$. No emission is detected from its counter-image, but since the lens model is robust we give this lens a grade A.

\item SL2SJ220629+005728. The image shows a secondary component with a color
similar to the main lens, in the proximity of one of the arcs. This component
could contribute to the overall lensing effect. We modeled it with a singular
isothermal sphere. The fit yielded a small value for its Einstein
radius as in Paper I.

%\item SL2SJ221045-005918. Only one arc is visible from the ground, leaving the lens model relatively unconstrained. The lens galaxy is very faint and the spectroscopic follow-up did not provide a redshift for both the source and the lens. The photometric redshift of this lens is $z_{\mathrm{phot}}=1.02$. {\bf Do we even want this lens here? True, it has a spectrum but that's junk.}

\item SL2SJ221326-000946 is a disky lens. A merging pair and a third image of the same bright
component are identified on the arc. No counter-image is visible in our
images.

\item SL2SJ221407-180712. Analogous to other systems with CFHT data
  only, we need to fix the PA of the mass distribution in order to constrain accurately the Einstein radius.
  It is a grade B because of the lack of source spectroscopy. 

\item SL2SJ221929-001743. Only one source component, at a spectroscopic redshift of $z=1.02$, is visible in the CFHT image. The constraints that this image provides on the lens model are not good enough and we need to fix the position angle of the mass to that of the light. The best-fit model does not predict multiple images. Grade B.

\item SL2SJ222012+010606. The CFHT image shows two blue components on opposite sides of the lens. The brighter arc is measured to be at a higher redshift than the lens, while we have no spectroscopic information on the fainter blob.
The lens model that we obtain is only partly satisfactory, because it predicts a mass centroid off by $\sim1.5$ pixels from the light centroid.
Moreover, the stellar mass and velocity dispersion of the foreground galaxy are unusually low in relation to the measured Einstein radius. It seems then plausible that the secondary source component is not a counter-image to the main arc.
The foreground galaxy is definitely providing some lensing, but probably not strong. Grade C.

\item SL2SJ222148+011542. Two arcs are visible both in photometry and in spectroscopy, making this a grade A lens.

\item SL2SJ222217+001202. An arc with no clear counter-image is visible in the
ground-based image of this lens. We put a Gaussian prior on the lens PA,
centered on the light PA and with a 10 degree dispersion, in order to obtain a
meaningful model of this lens. Grade B.

\end{itemize}

% %%%%%%%%%%%%%%%%%%%%%%%%%%%%%%%%%%%%%
\begin{deluxetable*}{cccccccc}
\tablewidth{0pt}
\tablecaption{Lens model parameters\label{table:lensfit}}
\tablehead{
\colhead{Name} & \colhead{$R_\mathrm{Ein}$} & \colhead{$q$} & \colhead{PA} &
\colhead{$m_{\rm{s}}$} & Grade & Notes & \hst? \\
& (arcsec) & & (degrees) & (mag) & & & }
\startdata
SL2SJ020833-071414 & $2.66 \pm 0.08$ & $0.76 \pm 0.01$ & $59.2 \pm  0.3$ & $24.88$ & B &  & Y \\ 
SL2SJ021206-075528 & $1.24 \pm 0.04$ & $0.77 \pm 0.02$ & $-12.7 \pm  3.1$ & $24.72$ & B &  & N \\ 
SL2SJ021247-055552 & $1.27 \pm 0.04$ & $0.82 \pm 0.04$ & $-34.3 \pm  3.5$ & $25.11$ & A &  & N \\ 
SL2SJ021325-074355 & $2.39 \pm 0.07$ & $0.54 \pm 0.01$ & $17.8 \pm  0.4$ & $23.68$ & A &  & Y \\ 
comp. 2 & $0.74\pm0.11$ & $0.34\pm0.34$ & 53.6 (fixed) & & & \\ 
SL2SJ021411-040502 & $1.41 \pm 0.04$ & $0.59 \pm 0.02$ & $-84.0 \pm  1.1$ & $24.61$ & A &  & Y \\ 
SL2SJ021737-051329 & $1.27 \pm 0.04$ & $0.85 \pm 0.02$ & $-75.1 \pm  3.3$ & $24.06$ & A &  & Y \\ 
$\gamma_{\mathrm{ext}}$ & $0.11\pm0.01$ & & $ 1.0\pm 0.1$ & & & \\ 
SL2SJ021801-080247 & $1.00 \pm 0.03$ & $0.81 \pm 0.03$ & $39.9 \pm  2.7$ & $24.79$ & B &  & N \\ 
SL2SJ021902-082934 & $1.30 \pm 0.04$ & $0.80 \pm 0.06$ & $-82.4 \pm  3.9$ & $26.31$ & A &  & Y \\ 
SL2SJ022046-094927 & $1.00 \pm 0.03$ & $0.96 \pm 0.02$ & $-38.0 \pm  9.8$ & $24.15$ & B &  & N \\ 
SL2SJ022056-063934 & $1.19 \pm 0.04$ & $0.61 \pm 0.04$ & $-79.8 \pm  2.5$ & $24.57$ & B &  & N \\ 
SL2SJ022346-053418 & $1.22 \pm 0.11$ & $0.39 \pm 0.09$ & $70.5 \pm  5.0$ & $24.35$ & B &  & N \\ 
SL2SJ022357-065142 & $1.36 \pm 0.04$ & $0.79 \pm 0.03$ & $66.7 \pm  2.6$ & $24.73$ & A &  & Y \\ 
SL2SJ022511-045433 & $1.76 \pm 0.05$ & $0.58 \pm 0.02$ & $24.6 \pm  0.4$ & $23.61$ & A &  & Y \\ 
SL2SJ022610-042011 & $1.19 \pm 0.04$ & $0.79 \pm 0.03$ & $-10.1 \pm  6.0$ & $25.26$ & A &  & Y \\ 
SL2SJ022648-040610 & $1.29 \pm 0.04$ & $0.79 \pm 0.07$ & $-64.3 \pm  7.5$ & $25.93$ & A & disky & Y \\ 
SL2SJ022648-090421 & $1.56 \pm 0.05$ & $0.87 \pm 0.03$ & $73.6 \pm  4.2$ & $26.16$ & B &  & N \\ 
SL2SJ023251-040823 & $1.04 \pm 0.03$ & $0.94 \pm 0.03$ & $76.8 \pm -60.9$ & $24.36$ & A &  & Y \\ 
SL2SJ084847-035103 & $0.85 \pm 0.07$ & $0.77 \pm 0.18$ & $80.5 \pm -57.5$ & $23.83$ & A &  & N \\ 
SL2SJ084909-041226 & $1.10 \pm 0.03$ & $0.73 \pm 0.03$ & $40.5 \pm  1.6$ & $24.16$ & A &  & Y \\ 
SL2SJ084934-043352 & $1.23 \pm 0.05$ & $0.63 \pm 0.12$ & $36.4$ (fixed) & $23.87$ & B &  & N \\ 
SL2SJ084959-025142 & $1.16 \pm 0.04$ & $0.72 \pm 0.04$ & $-62.7 \pm  2.3$ & $25.85$ & A &  & Y \\ 
SL2SJ085019-034710 & $0.93 \pm 0.03$ & $0.44 \pm 0.05$ & $ 2.2 \pm  3.7$ & $25.59$ & A & disky & N \\ 
SL2SJ085327-023745 & $1.31 \pm 0.04$ & $0.72 \pm 0.02$ & $-0.4 \pm  1.2$ & $23.07$ & A &  & N \\ 
SL2SJ085540-014730 & $1.03 \pm 0.04$ & $0.96 \pm 0.03$ & $-64.4 \pm 40.5$ & $25.32$ & A &  & N \\ 
SL2SJ085559-040917 & $1.36 \pm 0.10$ & $0.31 \pm 0.13$ & $39.4 \pm 10.0$ & $24.14$ & B &  & N \\ 
SL2SJ085826-014300 & $0.90 \pm 0.03$ & $0.91 \pm 0.04$ & $65.0 \pm 20.3$ & $26.36$ & A &  & Y \\ 
SL2SJ090106-025906 & $1.03 \pm 0.03$ & $0.45 \pm 0.02$ & $-19.2 \pm  1.2$ & $25.68$ & A &  & Y \\ 
SL2SJ090407-005952 & $1.40 \pm 0.04$ & $0.64 \pm 0.01$ & $71.8 \pm  0.8$ & $24.32$ & A &  & Y \\ 
SL2SJ095921+020638 & $0.74 \pm 0.02$ & $0.95 \pm 0.01$ & $-72.0 \pm  2.3$ & $26.79$ & A &  & Y \\ 
SL2SJ135847+545913 & $1.21 \pm 0.04$ & $0.76 \pm 0.01$ & $-71.1 \pm  1.6$ & $24.30$ & A &  & Y \\ 
SL2SJ135949+553550 & $1.14 \pm 0.03$ & $0.60 \pm 0.01$ & $55.7 \pm  0.5$ & $25.53$ & A &  & Y \\ 
SL2SJ140123+555705 & $1.74 \pm 0.07$ & $0.49 \pm 0.04$ & $-43.8 \pm  0.7$ & $26.63$ & A &  & Y \\ 
SL2SJ140156+554446 & $2.03 \pm 0.06$ & $0.58 \pm 0.01$ & $32.1 \pm  0.3$ & $24.41$ & A &  & Y \\ 
SL2SJ140221+550534 & $1.23 \pm 0.04$ & $0.75 \pm 0.04$ & $-48.2 \pm  3.0$ & $25.41$ & A &  & Y \\ 
SL2SJ140454+520024 & $2.55 \pm 0.08$ & $0.55 \pm 0.03$ & $70.9 \pm  0.8$ & $24.79$ & A &  & N \\ 
SL2SJ140533+550231 & $1.05 \pm 0.07$ & $0.55 \pm 0.02$ & $ 1.7 \pm  2.3$ & $24.15$ & A & double & Y \\ 
comp. 2 & $0.52\pm0.06$ & $0.81\pm0.06$ & $42.1\pm19.8$ & & & \\ 
SL2SJ140546+524311 & $1.51 \pm 0.05$ & $0.45 \pm 0.04$ & $-53.3 \pm  2.3$ & $26.44$ & A &  & Y \\ 
$\gamma_{\mathrm{ext}}$ & $0.05\pm0.02$ & & $-35.9\pm 0.1$ & & & \\ 
SL2SJ140614+520253 & $4.36 \pm 0.16$ & $0.64 \pm 0.03$ & $-52.5 \pm  1.8$ & $22.58$ & C &  & N \\ 
SL2SJ140650+522619 & $0.94 \pm 0.03$ & $1.00 \pm 0.01$ & $-26.8 \pm  0.5$ & $24.08$ & A &  & Y \\ 
SL2SJ141137+565119 & $0.93 \pm 0.03$ & $0.85 \pm 0.01$ & $62.4 \pm  0.3$ & $24.53$ & A &  & Y \\ 
SL2SJ141917+511729 & $3.11 \pm 0.26$ & $0.64 \pm 0.12$ & $49.5 \pm  4.5$ & $24.83$ & A &  & Y \\ 
SL2SJ142003+523137 & $1.79 \pm 0.09$ & $0.35 \pm 0.03$ & $65.6 \pm  2.0$ & $24.70$ & A & disky & N \\ 
SL2SJ142031+525822 & $0.96 \pm 0.14$ & $0.99 \pm 0.12$ & $-80.4 \pm  5.3$ & $22.45$ & B &  & Y \\ 
SL2SJ142059+563007 & $1.40 \pm 0.04$ & $0.67 \pm 0.01$ & $-10.4 \pm  0.3$ & $25.17$ & A &  & Y \\ 
SL2SJ142321+572243 & $1.30 \pm 0.14$ & $0.32 \pm 0.03$ & $43.9 \pm  1.2$ & $31.72$ & A &  & Y \\ 
SL2SJ142731+551645 & $0.81 \pm 0.03$ & $0.49 \pm 0.02$ & $-63.1 \pm  0.9$ & $24.73$ & A & disky & Y \\ 
SL2SJ220329+020518 & $1.95 \pm 0.06$ & $0.45 \pm 0.01$ & $-31.8 \pm  0.2$ & $24.95$ & A &  & Y \\ 
SL2SJ220506+014703 & $1.66 \pm 0.06$ & $0.74 \pm 0.04$ & $81.1 \pm  3.4$ & $23.68$ & A &  & N \\ 
SL2SJ220629+005728 & $1.55 \pm 0.07$ & $0.67 \pm 0.04$ & $-28.9 \pm  3.8$ & $24.78$ & B &  & N \\ 
comp. 2 & $0.16\pm0.06$ & & & & & & \\ 
SL2SJ221326-000946 & $1.07 \pm 0.03$ & $0.20 \pm 0.01$ & $-41.5 \pm  0.5$ & $24.82$ & A & disky & Y \\ 
SL2SJ221407-180712 & $1.06 \pm 0.22$ & $0.99 \pm 0.09$ & $57.0$ (fixed) & $25.01$ & B &  & N \\ 
SL2SJ221852+014038 & $1.38 \pm 0.08$ & $0.40 \pm 0.09$ & $-67.2 \pm  4.9$ & $25.54$ & B &  & N \\ 
SL2SJ221929-001743 & $0.52 \pm 0.13$ & $0.83 \pm 0.30$ & $85.1$ (fixed) & $23.46$ & B &  & N \\ 
SL2SJ222012+010606 & $2.16 \pm 0.07$ & $0.69 \pm 0.03$ & $-26.1 \pm  3.0$ & $24.09$ & C &  & N \\ 
SL2SJ222148+011542 & $1.40 \pm 0.05$ & $0.88 \pm 0.03$ & $77.7 \pm  3.8$ & $24.48$ & A &  & N \\ 
SL2SJ222217+001202 & $1.44 \pm 0.15$ & $0.82 \pm 0.30$ & $41.0 \pm  6.8$ & $24.83$ & B &  & N \\ 

\enddata
\tablecomments{Peak value and $68\%$ confidence interval of the posterior
probability distribution of each lens parameter, marginalized over the other
parameters.  Columns 2--4 correspond to the Einstein radius ($R_{\rm Ein}$), the axis ratio of the elliptical isodensity contours ($q$), and the position angle measured east of north (PA) of the SIE lens model.  Column 5 shows the magnitude of the de-lensed source in the band
used for the lensing analysis: the bluest available band for \hst\ data, or $g$
band for CFHT data. The typical uncertainty on the source magnitude is $\sim0.5$. Column 6 lists notes on the lens morphology, while column 7 indicates whether the lens has \hst\ imaging.}
\end{deluxetable*}
% %%%%%%%%%%%%%%%%%%%%%%%%%%%%%%%%%%%%%

% %%%%%%%%%%%%%%%%%%%%%%%%%%%%%%%%%%%%%%%%
% MASSIVE MULTI-PAGE FIGURE OF LENS+MODELS
% %%%%%%%%%%%%%%%%%%%%%%%%%%%%%%%%%%%%%%%%
%latex code for the lens model figures

\def\fs{0.9} 

\begin{figure*}[!]
 \begin{center}
   \includegraphics[width=\fs\textwidth]{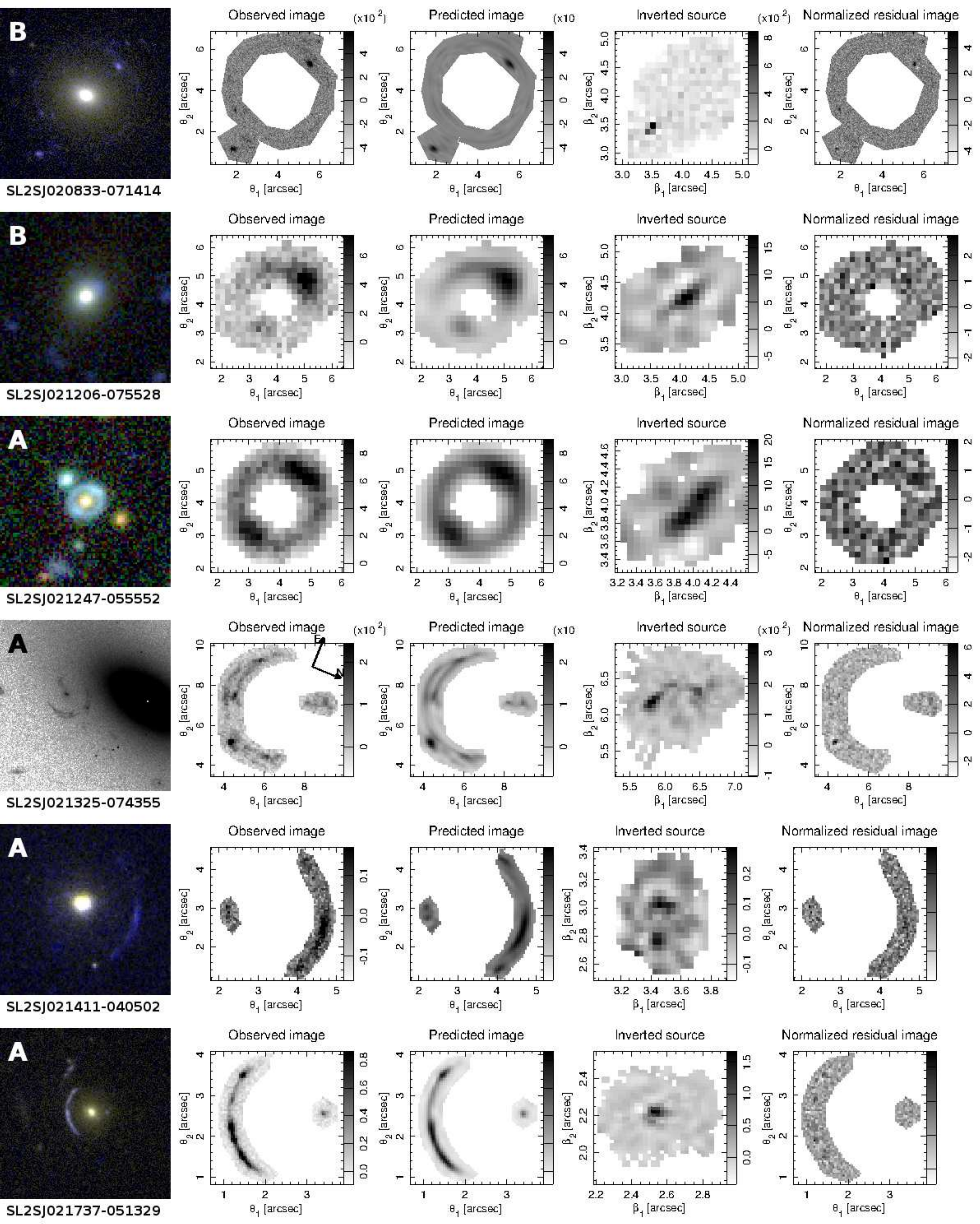} \\
 \end{center}
\caption{\label{fig:lenscuts} Lens modeling results showing, on each
  row, from left to right, a color cutout image, the input science imaged used
  for the modeling with uninteresting areas cropped out, the
  reconstructed lensed image, the reconstructed intrinsic source and
  the difference image (data$-$model) normalized in units of the estimated pixel uncertainties.}
\end{figure*}
 
\begin{figure*}[!]
 \begin{center}
   \includegraphics[width=\fs\textwidth]{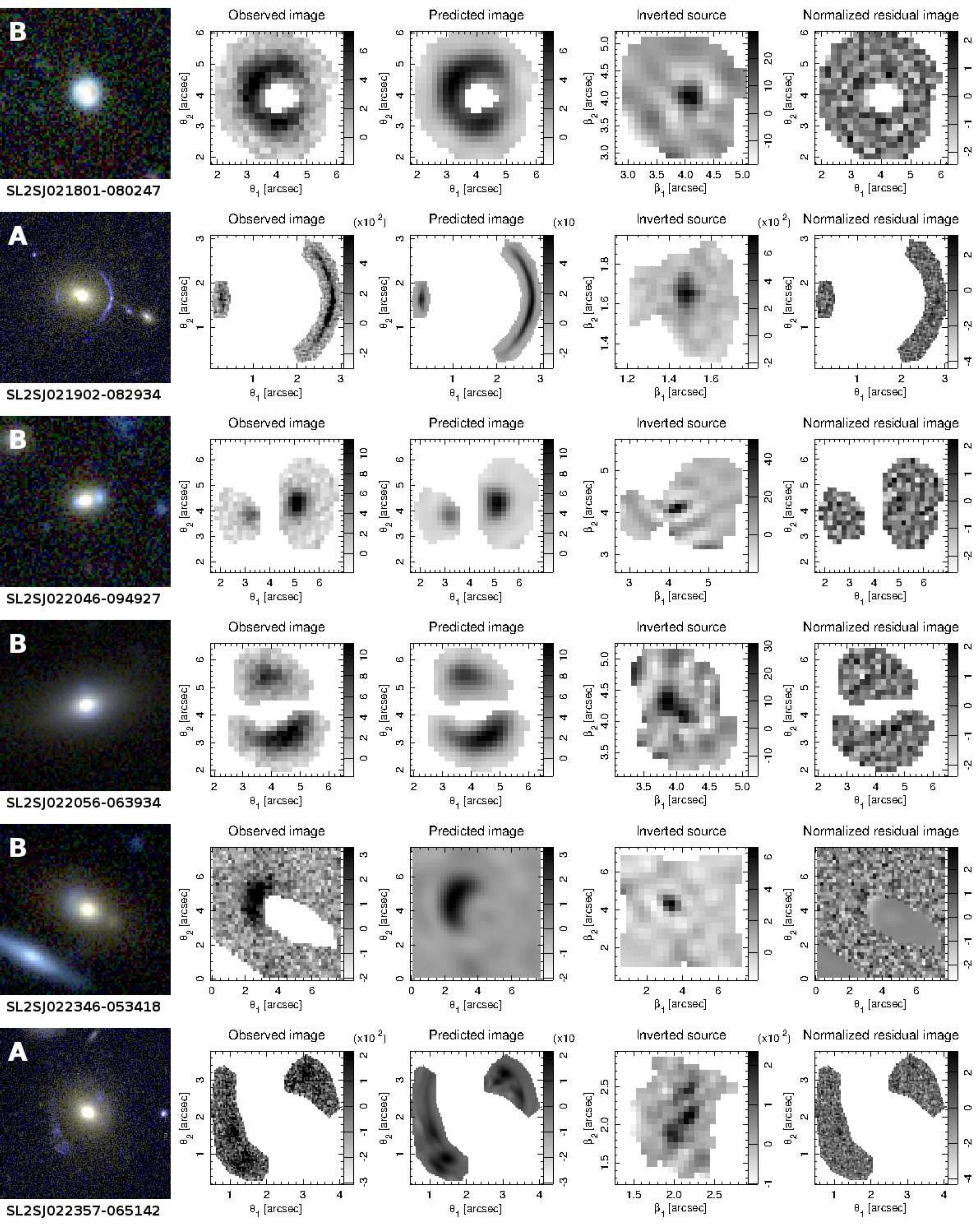} \\
 \end{center}
\figurenum{\ref{fig:lenscuts}}
\caption{continued.}
\end{figure*}
 
\begin{figure*}[!]
 \begin{center}
   \includegraphics[width=\fs\textwidth]{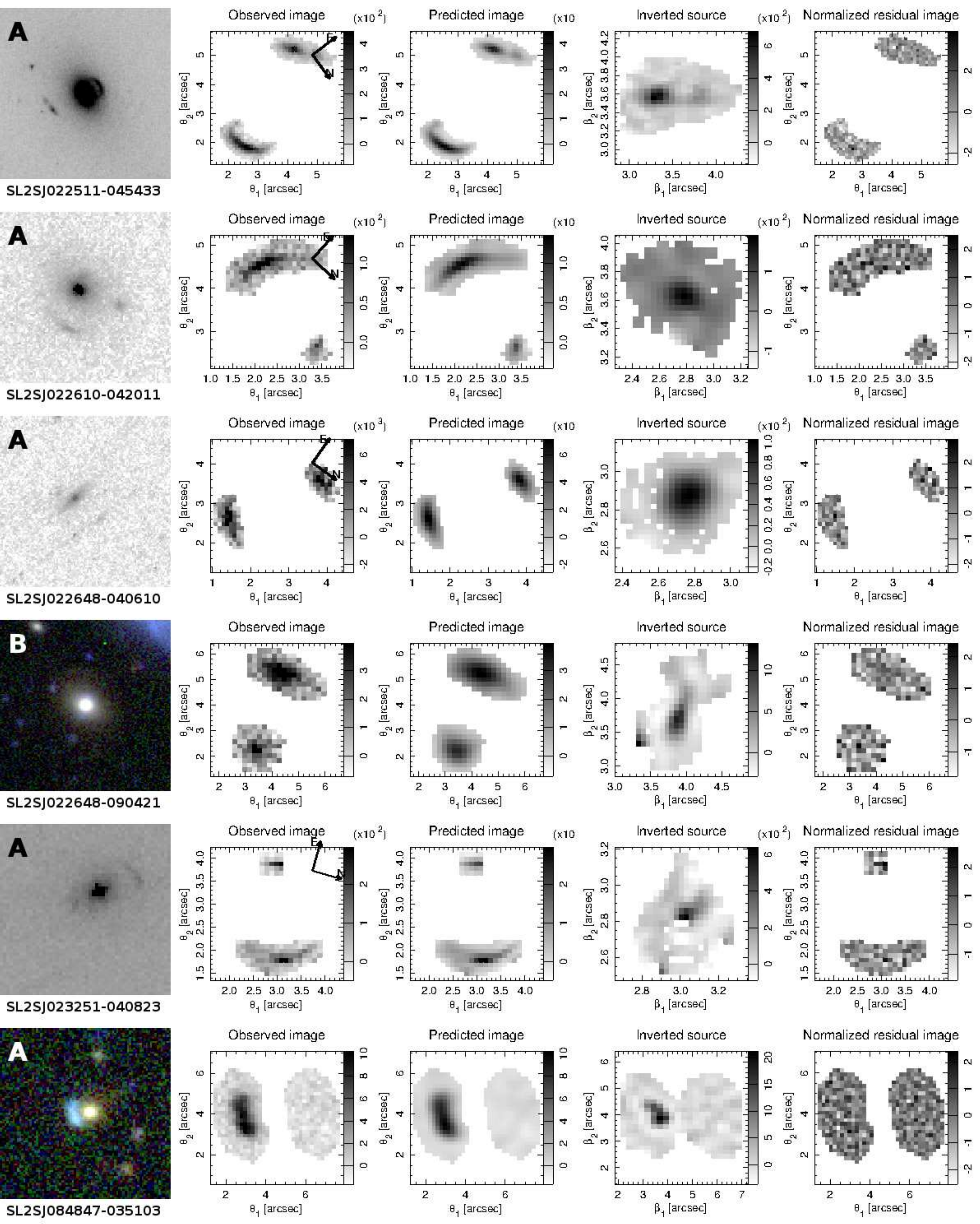} \\
 \end{center}
\figurenum{\ref{fig:lenscuts}}
\caption{continued.}
\end{figure*}
 
\begin{figure*}[!]
 \begin{center}
   \includegraphics[width=\fs\textwidth]{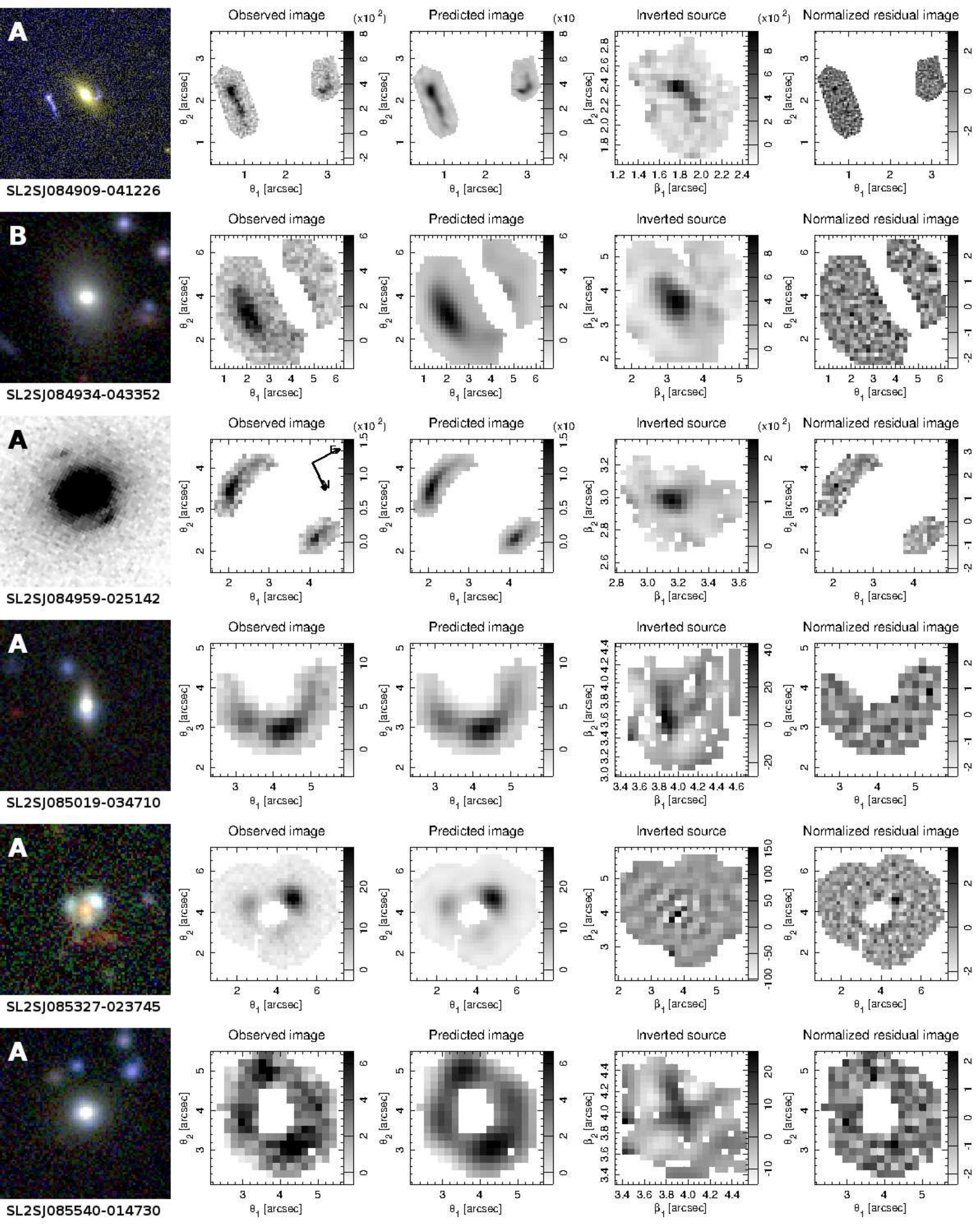} \\
 \end{center}
\figurenum{\ref{fig:lenscuts}}
\caption{continued.}
\end{figure*}
 
\begin{figure*}[!]
 \begin{center}
   \includegraphics[width=\fs\textwidth]{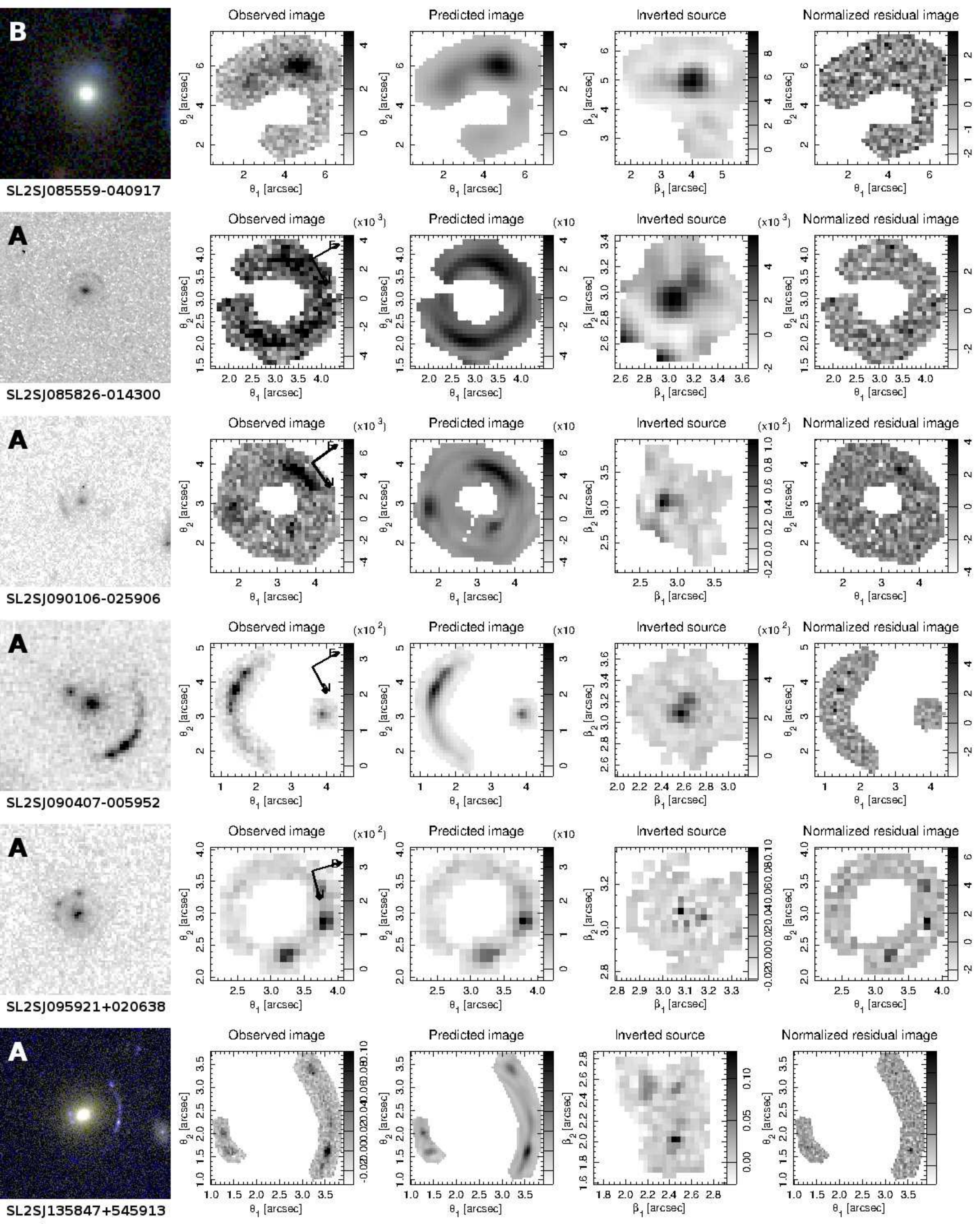} \\
 \end{center}
\figurenum{\ref{fig:lenscuts}}
\caption{continued.}
\end{figure*}
 
\begin{figure*}[!]
 \begin{center}
   \includegraphics[width=\fs\textwidth]{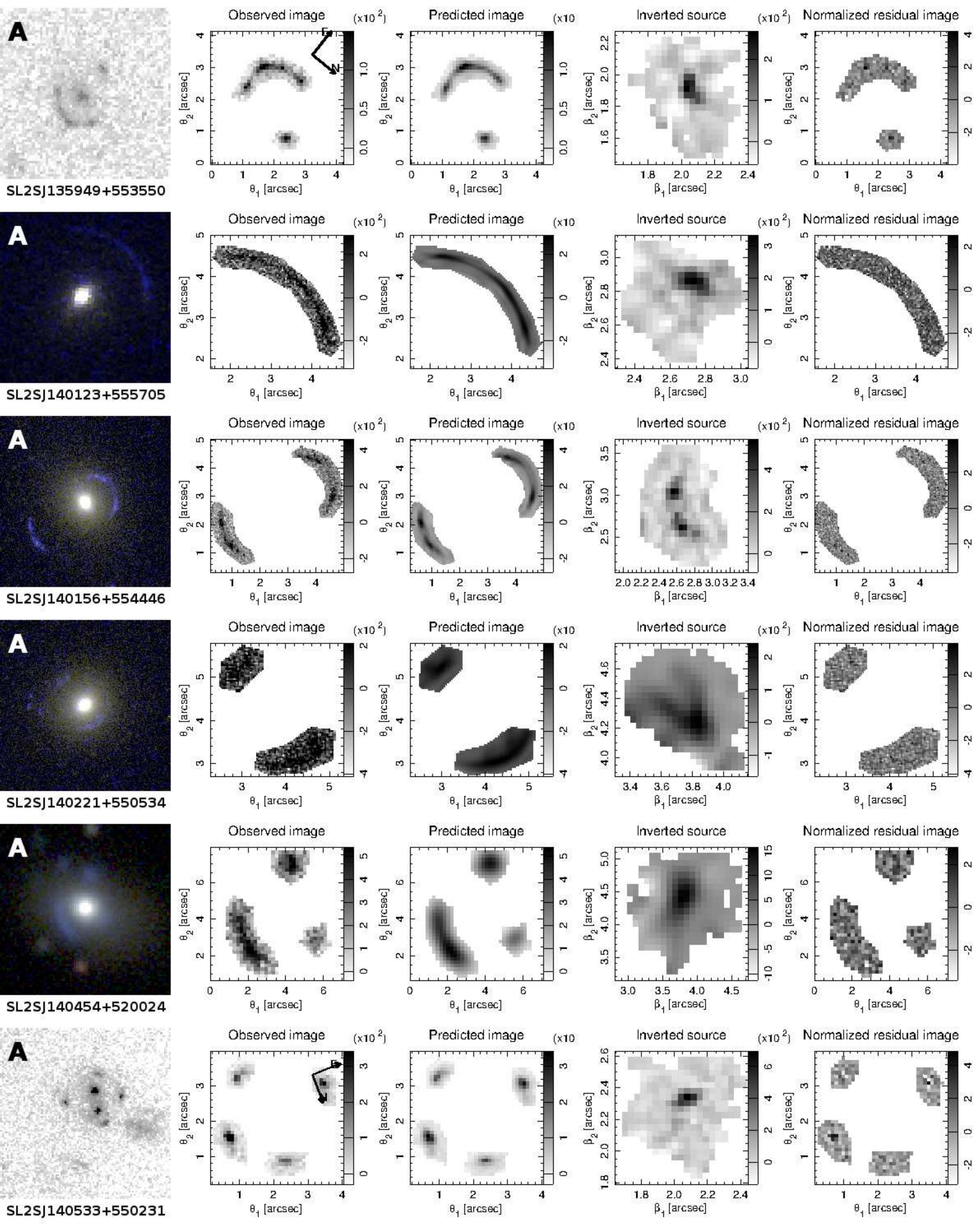} \\
 \end{center}
\figurenum{\ref{fig:lenscuts}}
\caption{continued.}
\end{figure*}
 
\begin{figure*}[!]
 \begin{center}
   \includegraphics[width=\fs\textwidth]{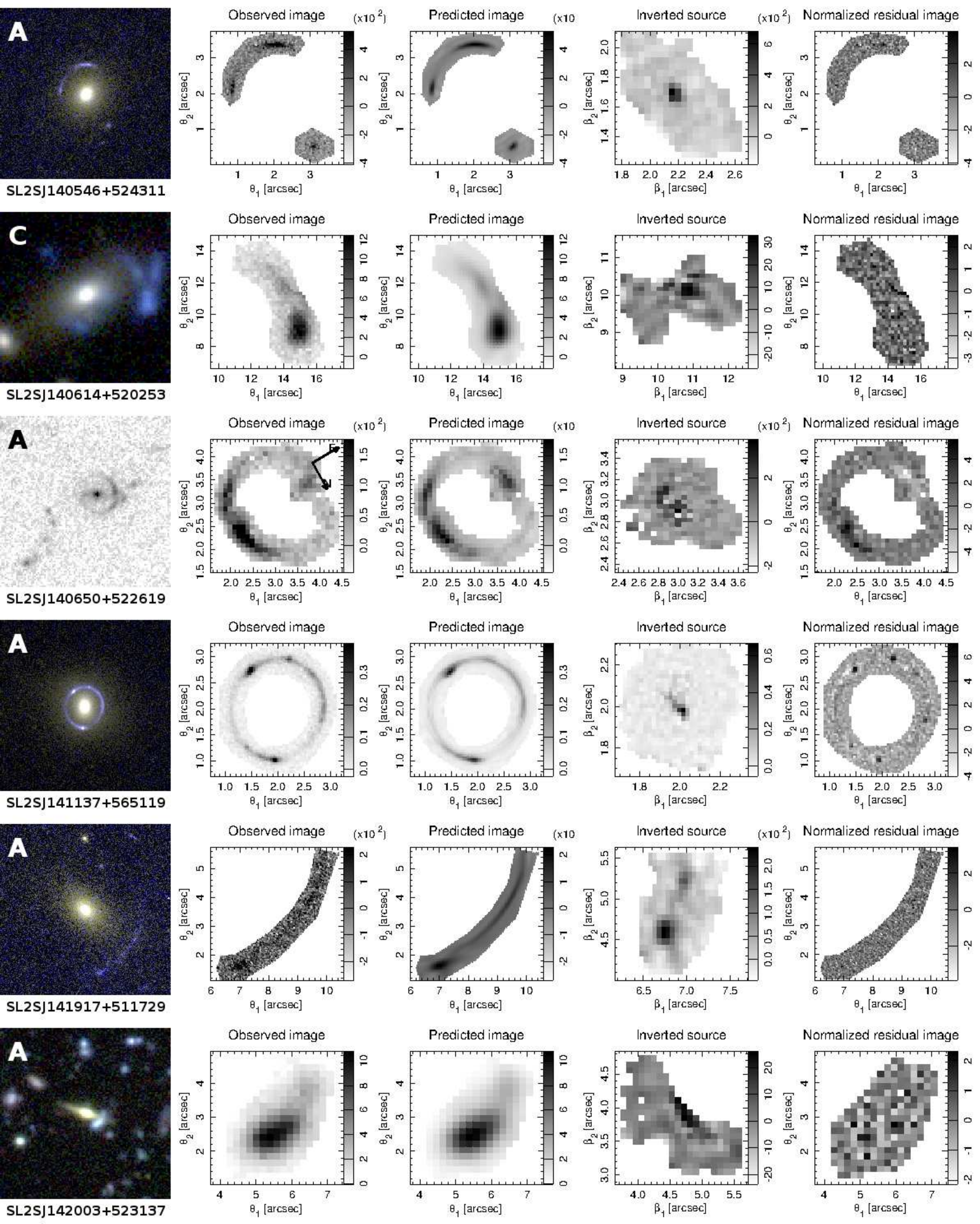} \\
 \end{center}
\figurenum{\ref{fig:lenscuts}}
\caption{continued.}
\end{figure*}
 
\begin{figure*}[!]
 \begin{center}
   \includegraphics[width=\fs\textwidth]{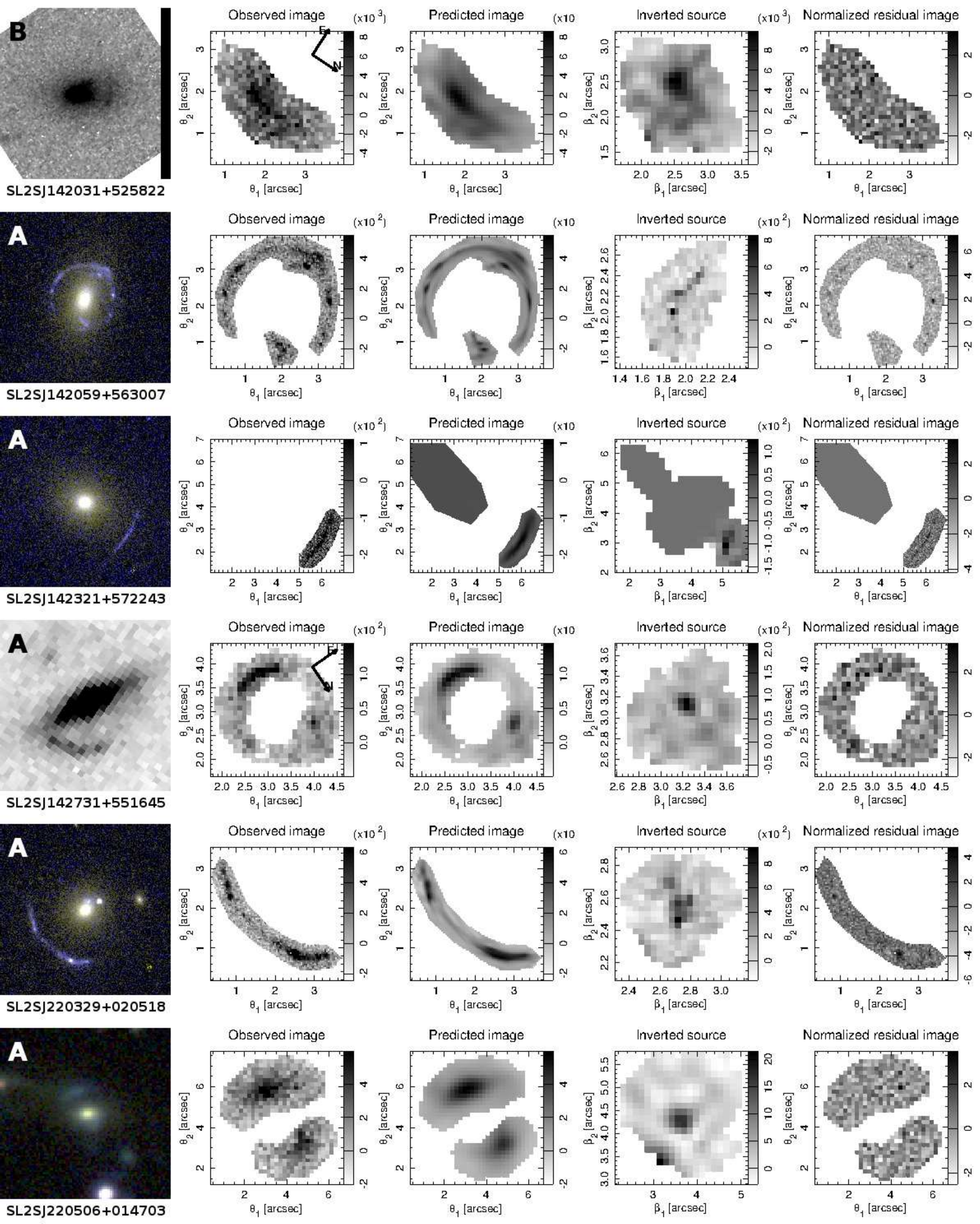} \\
 \end{center}
\figurenum{\ref{fig:lenscuts}}
\caption{continued.}
\end{figure*}
 
\begin{figure*}[!]
 \begin{center}
   \includegraphics[width=\fs\textwidth]{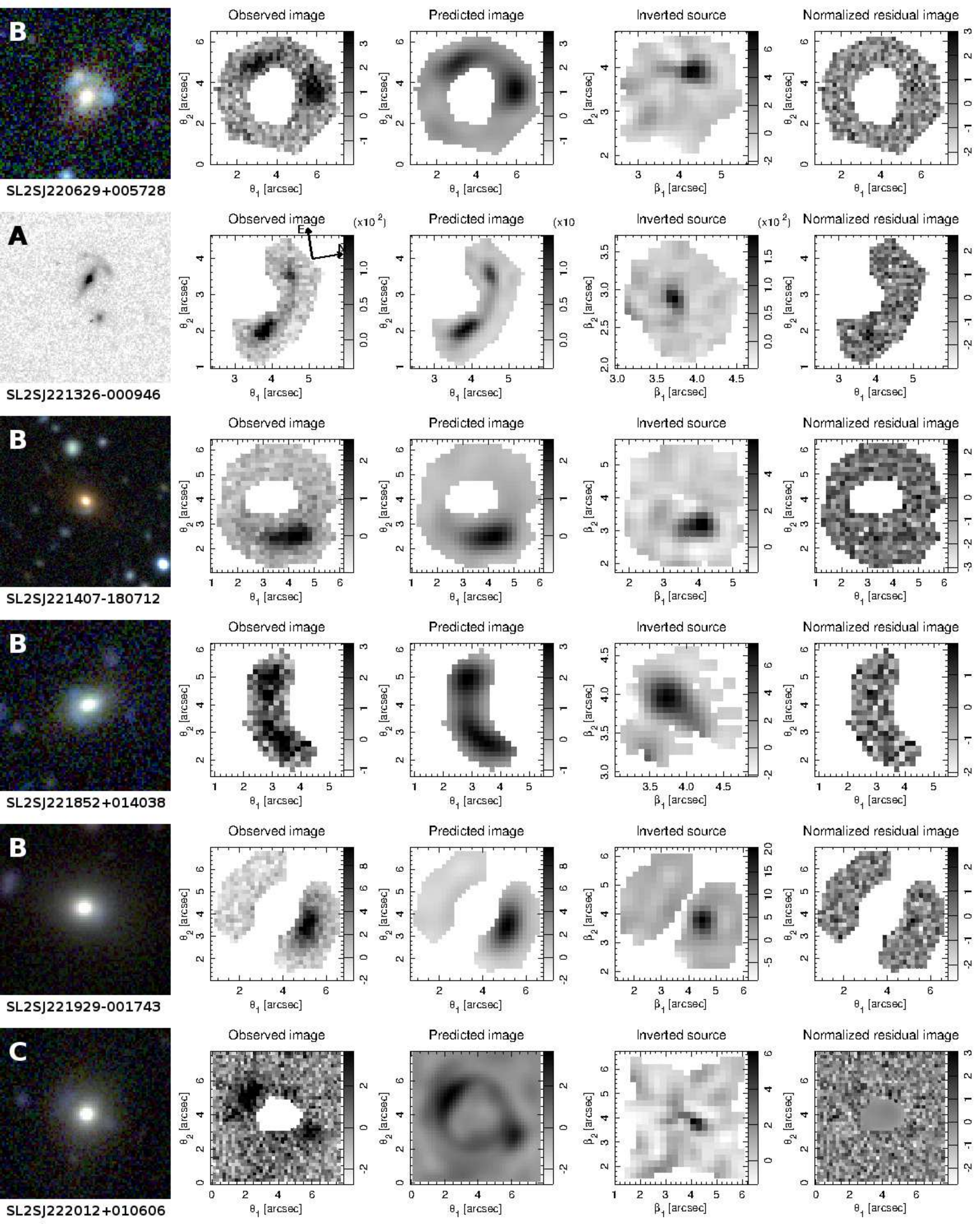} \\
 \end{center}
\figurenum{\ref{fig:lenscuts}}
\caption{continued.}
\end{figure*}
 
\begin{figure*}[!]
 \begin{center}
   \includegraphics[width=\fs\textwidth]{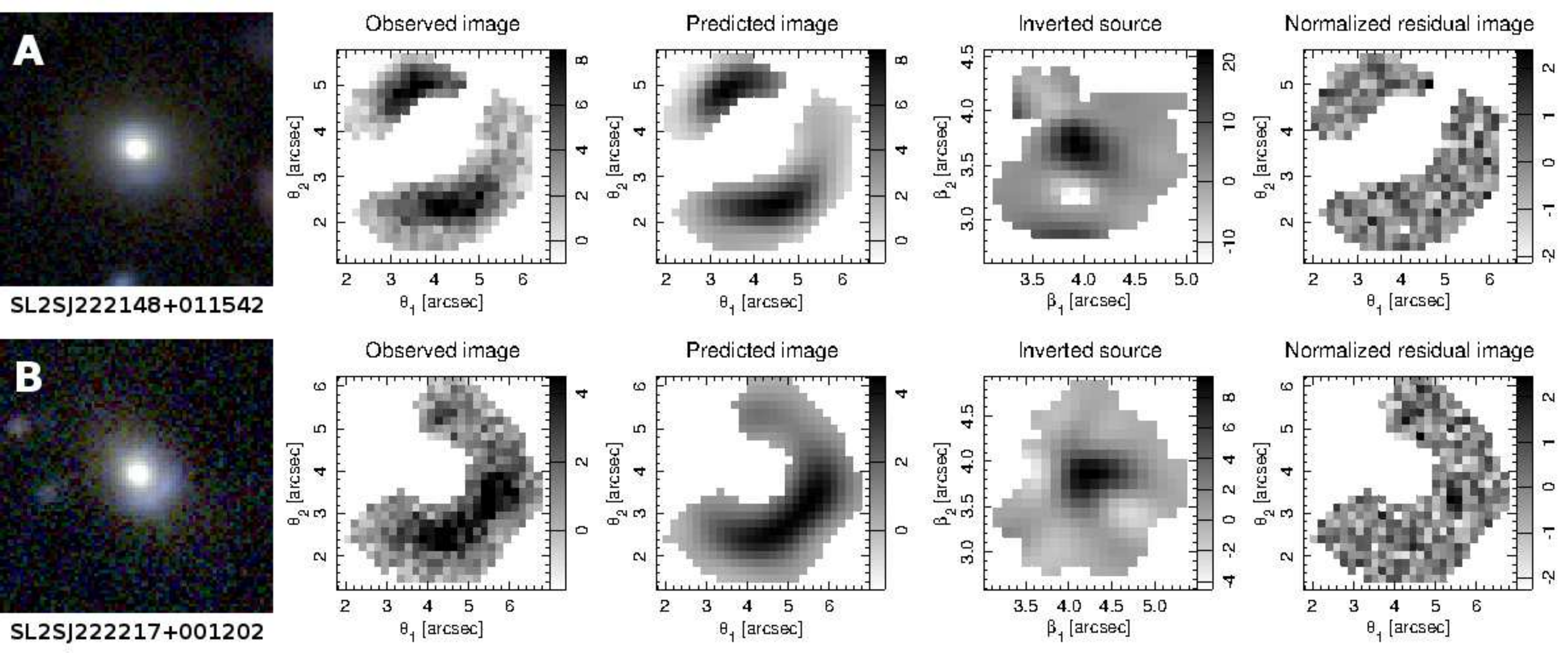} \\
 \end{center}
\figurenum{\ref{fig:lenscuts}}
\caption{continued.}
\end{figure*}

% %%%%%%%%%%%%%%%%%%%%%%%%%%%%%%%%%%%%%%%

%-------------------------------------------------------------------------------

\section{Stellar masses}\label{sect:mstar}

One of the goals of our study is to better understand the mass assembly of early-type galaxies over cosmic time.
While gravitational lensing provides us with a precise measurement of the total mass enclosed within the Einstein radius of our lenses, measurements of the stellar mass are needed to separate the contribution of baryonic and dark matter to the total mass balance.
In this paper we estimate stellar masses through stellar population synthesis (SPS) fitting of our photometric measurements: we create stellar populations assuming a simply-parametrized star formation history and stellar initial mass function (IMF), calculate magnitudes in the observed bands and fit to the measurements.
The implementation of this procedure is the same as the one in \citet{Aug++09} and is based on a code written by M. W. Auger. 
We create composite stellar populations from stellar templates by \citet{B+C03}, with both a Salpeter and a Chabrier IMF. We assume an exponentially declining star formation history, appropriate given the old age of the red galaxies in our sample.
%, and adopt the mass-metallicity relation from \citet{Gal++05} as a prior on these two parameters {\bf It turns out it doesn't make any difference on M*.}
In order to obtain robust stellar masses, measurements in a few different bands are needed.
Although \hst\ images provide better spatial resolution, useful to deblend the lens light from that of the background source, our objects have \hst\ data in at most two bands which are not enough for the purpose of fitting SPS models.
CFHT images on the other hand are deep and available consistently in five different bands for all of the targets.
The inclusion of the \hst\ photometry to the overall SED fitting would
not bring much new information and we therefore discard it. %Therefore we use the latter data for the fit.
The fit is based on an MCMC sampling.
The measured values of the stellar masses are reported in \Tref{table:mstar}.

\begin{figure}
\includegraphics[width=\columnwidth]{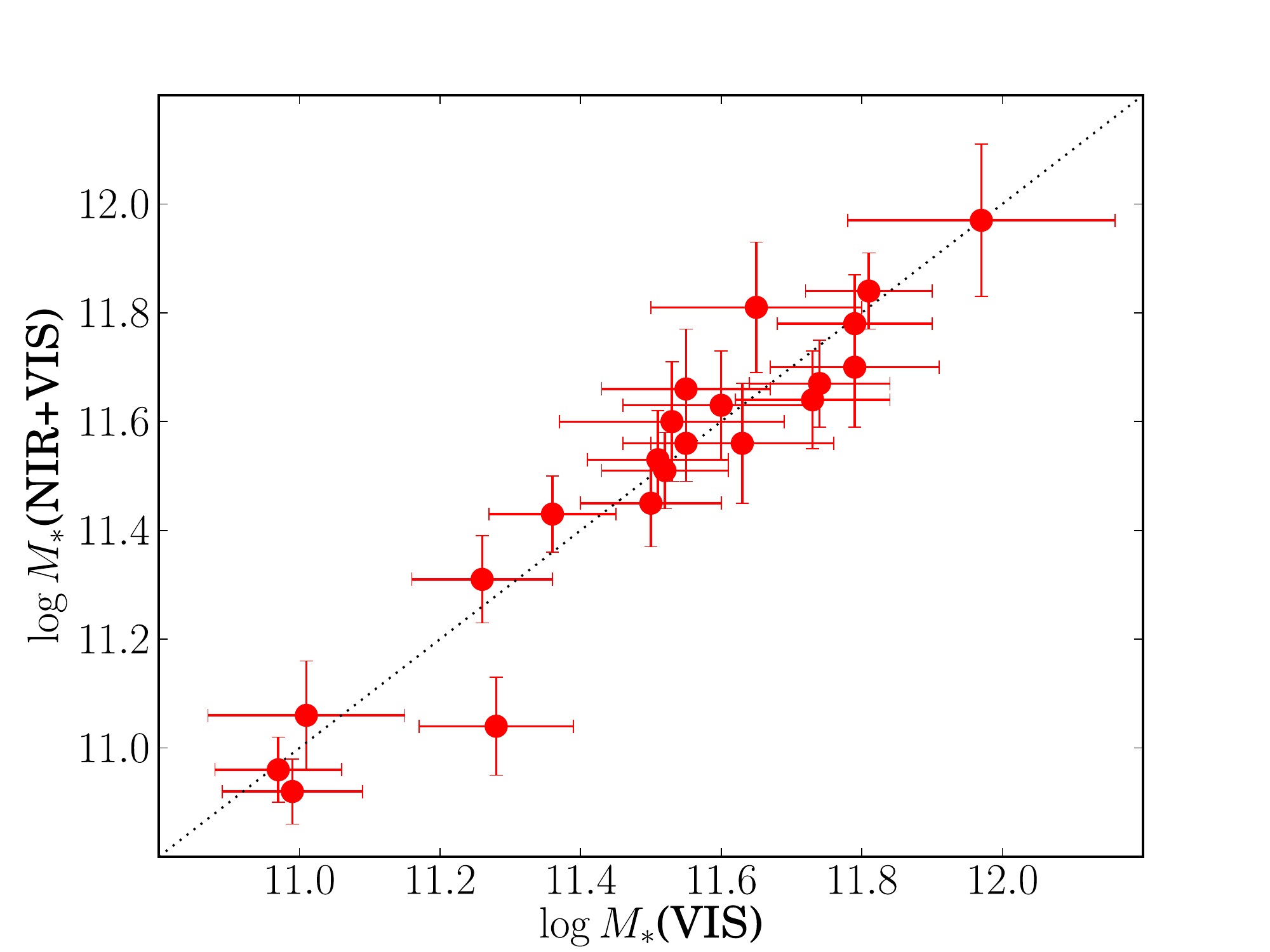}
\caption{\label{fig:mstar}. Comparison of stellar masses obtained with
either optical $ugriz$ bands only or with optical + near IR
bands, for a Salpeter IMF. We observe no 
significant differences in the recovered masses.}
\end{figure}

For the systems with additional NIR observations the fit is repeated including those data.
The addition of NIR fluxes produces stellar masses consistent with the values measured with optical data only, but with smaller uncertainty (see \Fref{fig:mstar}).
The relative scatter between stellar masses obtained from optical photometry alone and with the addition of NIR data is $0.06$ dex in $\log{M_*}$ and the bias is $0.01$. This gives us an estimate of the systematic error coming from the stellar templates being not a perfect description of the data over all photometric bands; in Paper IV, this systematic uncertainty is added to the statistical uncertainty on $M_*$ when dealing with stellar masses.
On the one hand the tight agreement between optical and optical+NIR stellar masses should not come as a surprise since the two data sets differ in most cases only by the addition of one band.
On the other hand, if the optical data were contaminated with poor subtraction of light from the blue arcs the resulting stellar masses could be biased. The fact that NIR data, with little to no contamination from the background source, does not change the inference is reassuring on the quality of our photometric measurements.

Some of the stellar masses measured here are not consistent with previous measurements from Paper II. This reflects the difference in the measured magnitudes due to the different source masking strategy discussed in \Sref{ssec:photolenses}.
The values reported here are to be considered more robust.

\begin{figure}
\includegraphics[width=\columnwidth]{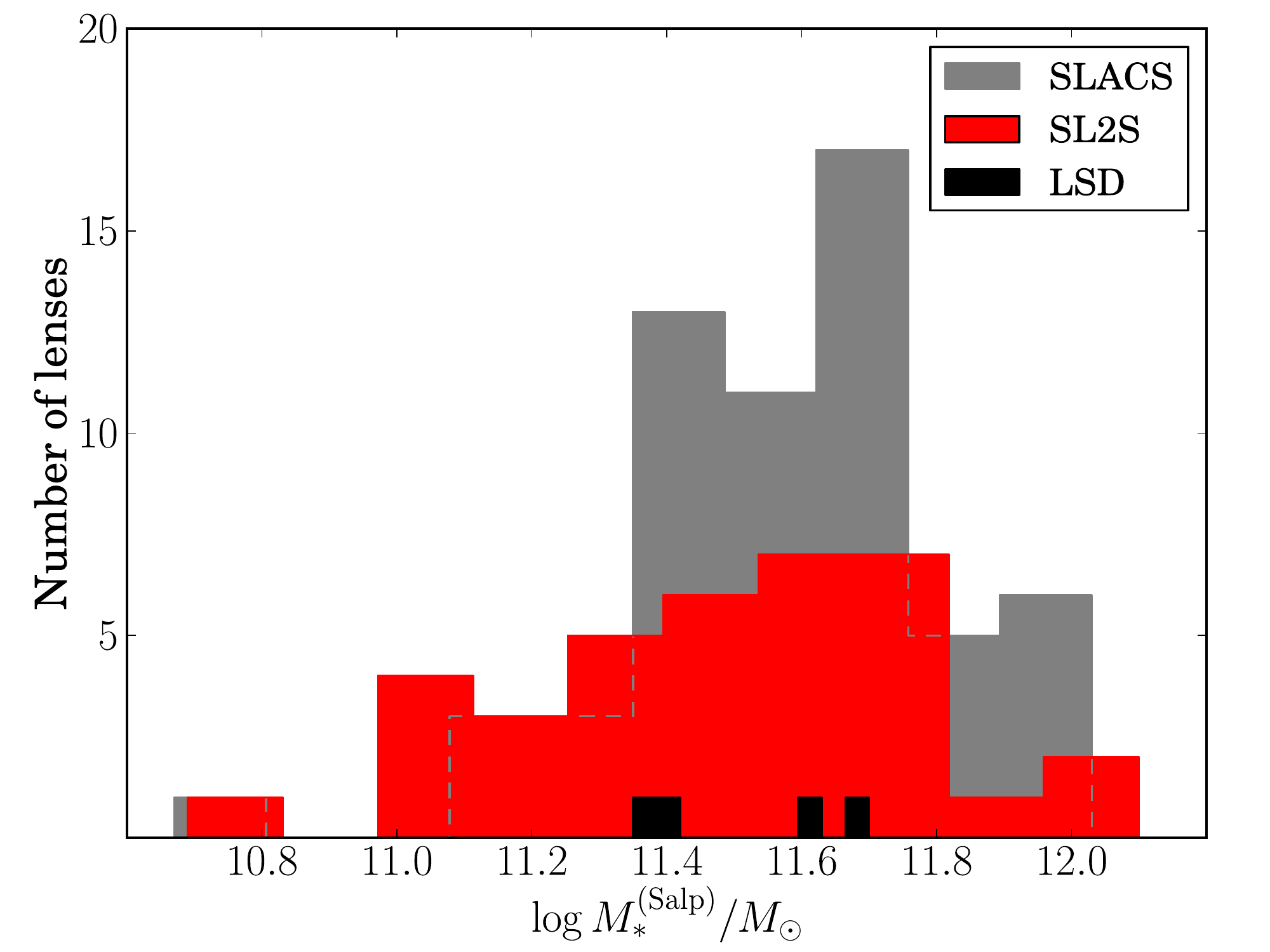}
\caption{\label{fig:mstarhist} Distribution in stellar mass of the grade A SL2S, SLACS and LSD lenses. SLACS stellar masses are from \citet{Aug++10} and LSD masses are taken from \citet{Ruf++11}. Stellar masses are obtained assuming a Salpeter IMF.
}
\end{figure}

The median stellar mass of the sub-sample of grade A SL2S lenses is $10^{11.53}M_\odot$, if a Salpeter IMF is assumed, and the standard deviation of the sample is 0.3 dex in $\log{M_*}$.
The distribution in stellar mass of SL2S galaxies is very similar to that of SLACS galaxies, as shown in \Fref{fig:mstarhist}.
This is important in view of analyses that combine data from both samples, as we do in Paper IV. 

% %%%%%%%%%%%%%%%%%%%%%%%%%%%%%%%%%%%%%
\begin{deluxetable*}{cccccc}
\tablewidth{0pt}
\tablecaption{Stellar mass measurements}
\tablehead{
\colhead{Lens name} & \colhead{$z$} & \colhead{$\log{M_*^{(\rm{Chab})}/M_\odot}$} & \colhead{$\log{M_*^{(\rm{Chab})}(\rm{NIR})/M_\odot}$} &\colhead{$\log{M_*^{(\rm{Salp})}/M_\odot}$} & \colhead{$\log{M_*^{(\rm{Salp})}(\rm{NIR})/M_\odot}$} }
\startdata
SL2SJ020833-071414 & $0.428$ & $11.59 \pm 0.10$ & $\cdots$ & $11.84 \pm 0.10$ & $\cdots$ \\ 
SL2SJ021206-075528 & $0.460$ & $11.33 \pm 0.10$ & $\cdots$ & $11.59 \pm 0.10$ & $\cdots$ \\ 
SL2SJ021247-055552 & $0.750$ & $11.17 \pm 0.17$ & $\cdots$ & $11.45 \pm 0.17$ & $\cdots$ \\ 
SL2SJ021325-074355 & $0.717$ & $11.71 \pm 0.18$ & $11.73 \pm 0.15$ & $11.97 \pm 0.19$ & $11.97 \pm 0.14$ \\ 
SL2SJ021411-040502 & $0.609$ & $11.34 \pm 0.14$ & $11.38 \pm 0.10$ & $11.60 \pm 0.14$ & $11.63 \pm 0.10$ \\ 
SL2SJ021737-051329 & $0.646$ & $11.29 \pm 0.15$ & $11.35 \pm 0.11$ & $11.53 \pm 0.16$ & $11.60 \pm 0.11$ \\ 
SL2SJ021801-080247 & $\cdots$ & $\cdots$ & $\cdots$ & $\cdots$ & $\cdots$ \\ 
SL2SJ021902-082934 & $0.389$ & $11.24 \pm 0.10$ & $11.20 \pm 0.08$ & $11.50 \pm 0.10$ & $11.45 \pm 0.08$ \\ 
SL2SJ022046-094927 & $0.572$ & $11.11 \pm 0.12$ & $\cdots$ & $11.36 \pm 0.11$ & $\cdots$ \\ 
SL2SJ022056-063934 & $0.330$ & $11.44 \pm 0.10$ & $\cdots$ & $11.69 \pm 0.09$ & $\cdots$ \\ 
SL2SJ022346-053418 & $0.499$ & $11.51 \pm 0.11$ & $\cdots$ & $11.76 \pm 0.11$ & $\cdots$ \\ 
SL2SJ022357-065142 & $0.473$ & $11.49 \pm 0.10$ & $11.44 \pm 0.08$ & $11.74 \pm 0.10$ & $11.67 \pm 0.08$ \\ 
SL2SJ022511-045433 & $0.238$ & $11.57 \pm 0.09$ & $11.59 \pm 0.07$ & $11.81 \pm 0.09$ & $11.84 \pm 0.07$ \\ 
SL2SJ022610-042011 & $0.494$ & $11.48 \pm 0.10$ & $11.41 \pm 0.09$ & $11.73 \pm 0.11$ & $11.64 \pm 0.09$ \\ 
SL2SJ022648-040610 & $0.766$ & $11.53 \pm 0.12$ & $11.46 \pm 0.11$ & $11.79 \pm 0.12$ & $11.70 \pm 0.11$ \\ 
SL2SJ022648-090421 & $0.456$ & $11.72 \pm 0.10$ & $\cdots$ & $11.97 \pm 0.10$ & $\cdots$ \\ 
SL2SJ023251-040823 & $0.352$ & $11.11 \pm 0.10$ & $11.18 \pm 0.08$ & $11.36 \pm 0.09$ & $11.43 \pm 0.07$ \\ 
SL2SJ084847-035103 & $0.682$ & $10.97 \pm 0.16$ & $\cdots$ & $11.24 \pm 0.16$ & $\cdots$ \\ 
SL2SJ084909-041226 & $0.722$ & $11.39 \pm 0.14$ & $11.31 \pm 0.10$ & $11.63 \pm 0.13$ & $11.56 \pm 0.11$ \\ 
SL2SJ084934-043352 & $0.373$ & $11.42 \pm 0.10$ & $\cdots$ & $11.67 \pm 0.10$ & $\cdots$ \\ 
SL2SJ084959-025142 & $0.274$ & $11.27 \pm 0.09$ & $11.27 \pm 0.07$ & $11.52 \pm 0.09$ & $11.51 \pm 0.07$ \\ 
SL2SJ085019-034710 & $0.337$ & $10.89 \pm 0.09$ & $\cdots$ & $11.14 \pm 0.09$ & $\cdots$ \\ 
SL2SJ085327-023745 & $0.774$ & $11.13 \pm 0.16$ & $\cdots$ & $11.38 \pm 0.16$ & $\cdots$ \\ 
SL2SJ085540-014730 & $0.365$ & $10.86 \pm 0.10$ & $\cdots$ & $11.11 \pm 0.10$ & $\cdots$ \\ 
SL2SJ085559-040917 & $0.419$ & $11.39 \pm 0.10$ & $\cdots$ & $11.63 \pm 0.10$ & $\cdots$ \\ 
SL2SJ085826-014300 & $0.580$ & $10.76 \pm 0.14$ & $10.81 \pm 0.10$ & $11.01 \pm 0.14$ & $11.06 \pm 0.10$ \\ 
SL2SJ090106-025906 & $0.670$ & $10.80 \pm 0.17$ & $\cdots$ & $11.07 \pm 0.16$ & $\cdots$ \\ 
SL2SJ090407-005952 & $0.611$ & $11.30 \pm 0.11$ & $11.41 \pm 0.11$ & $11.55 \pm 0.12$ & $11.66 \pm 0.11$ \\ 
SL2SJ095921+020638 & $0.552$ & $11.03 \pm 0.10$ & $10.81 \pm 0.09$ & $11.28 \pm 0.11$ & $11.04 \pm 0.09$ \\ 
SL2SJ135847+545913 & $0.510$ & $11.39 \pm 0.11$ & $\cdots$ & $11.66 \pm 0.11$ & $\cdots$ \\ 
SL2SJ135949+553550 & $0.783$ & $11.17 \pm 0.15$ & $\cdots$ & $11.41 \pm 0.15$ & $\cdots$ \\ 
SL2SJ140123+555705 & $0.527$ & $11.54 \pm 0.11$ & $\cdots$ & $11.80 \pm 0.11$ & $\cdots$ \\ 
SL2SJ140156+554446 & $0.464$ & $11.59 \pm 0.10$ & $\cdots$ & $11.85 \pm 0.10$ & $\cdots$ \\ 
SL2SJ140221+550534 & $0.412$ & $11.54 \pm 0.10$ & $\cdots$ & $11.79 \pm 0.10$ & $\cdots$ \\ 
SL2SJ140454+520024 & $0.456$ & $11.85 \pm 0.10$ & $\cdots$ & $12.10 \pm 0.10$ & $\cdots$ \\ 
SL2SJ140533+550231 & $\cdots$ & $\cdots$ & $\cdots$ & $\cdots$ & $\cdots$ \\ 
SL2SJ140546+524311 & $0.526$ & $11.42 \pm 0.11$ & $\cdots$ & $11.67 \pm 0.11$ & $\cdots$ \\ 
SL2SJ140614+520253 & $0.480$ & $11.68 \pm 0.11$ & $\cdots$ & $11.93 \pm 0.11$ & $\cdots$ \\ 
SL2SJ140650+522619 & $0.716$ & $11.34 \pm 0.15$ & $\cdots$ & $11.60 \pm 0.15$ & $\cdots$ \\ 
SL2SJ141137+565119 & $0.322$ & $11.04 \pm 0.09$ & $\cdots$ & $11.28 \pm 0.09$ & $\cdots$ \\ 
SL2SJ141917+511729 & $\cdots$ & $\cdots$ & $\cdots$ & $\cdots$ & $\cdots$ \\ 
SL2SJ142003+523137 & $0.354$ & $10.44 \pm 0.10$ & $\cdots$ & $10.69 \pm 0.10$ & $\cdots$ \\ 
SL2SJ142031+525822 & $0.380$ & $11.31 \pm 0.10$ & $\cdots$ & $11.56 \pm 0.09$ & $\cdots$ \\ 
SL2SJ142059+563007 & $0.483$ & $11.52 \pm 0.10$ & $\cdots$ & $11.76 \pm 0.10$ & $\cdots$ \\ 
SL2SJ142321+572243 & $\cdots$ & $\cdots$ & $\cdots$ & $\cdots$ & $\cdots$ \\ 
SL2SJ142731+551645 & $0.511$ & $10.97 \pm 0.12$ & $\cdots$ & $11.20 \pm 0.12$ & $\cdots$ \\ 
SL2SJ220329+020518 & $0.400$ & $11.00 \pm 0.09$ & $11.05 \pm 0.08$ & $11.26 \pm 0.10$ & $11.31 \pm 0.08$ \\ 
SL2SJ220506+014703 & $0.476$ & $11.26 \pm 0.11$ & $11.29 \pm 0.09$ & $11.51 \pm 0.10$ & $11.53 \pm 0.09$ \\ 
SL2SJ220629+005728 & $0.704$ & $11.40 \pm 0.15$ & $11.56 \pm 0.12$ & $11.65 \pm 0.15$ & $11.81 \pm 0.12$ \\ 
SL2SJ221326-000946 & $0.338$ & $10.73 \pm 0.09$ & $10.67 \pm 0.06$ & $10.99 \pm 0.10$ & $10.92 \pm 0.06$ \\ 
SL2SJ221407-180712 & $0.651$ & $\cdots$ & $\cdots$ & $\cdots$ & $\cdots$ \\ 
SL2SJ221852+014038 & $0.564$ & $11.52 \pm 0.11$ & $11.52 \pm 0.09$ & $11.79 \pm 0.11$ & $11.78 \pm 0.09$ \\ 
SL2SJ221929-001743 & $0.289$ & $11.32 \pm 0.09$ & $\cdots$ & $11.56 \pm 0.09$ & $\cdots$ \\ 
SL2SJ222012+010606 & $0.232$ & $10.73 \pm 0.10$ & $10.72 \pm 0.07$ & $10.97 \pm 0.09$ & $10.96 \pm 0.06$ \\ 
SL2SJ222148+011542 & $0.325$ & $11.30 \pm 0.09$ & $11.31 \pm 0.07$ & $11.55 \pm 0.09$ & $11.56 \pm 0.07$ \\ 
SL2SJ222217+001202 & $0.436$ & $11.26 \pm 0.10$ & $\cdots$ & $11.50 \pm 0.10$ & $\cdots$ \\ 

\enddata
\tablecomments{\label{table:mstar} Stellar masses from the fit of
  stellar population synthesis models to photometric data. The
  redshift of the lens galaxies is reported in column (2) and
  extensively discussed in Paper IV. %Stellar masses cannot be calculated for galaxies of unknown redshift.
}
\end{deluxetable*}
% %%%%%%%%%%%%%%%%%%%%%%%%%%%%%%%%%%%%%

%-------------------------------------------------------------------------------

\section{Summary and Conclusions}\label{sect:summary} 

We presented photometric measurements, lens models and stellar mass
measurements for a sample of \Nsys~systems, of which \NgradeA\ are grade A
(definite lenses) and 15 are grade B (probable lenses). We find that \hst\
imaging, even in snapshot mode, offers a clear-cut way to determine
whether SL2S candidates are actual lenses. Not surprisingly, most grade
A lenses are found for systems with \hst\ data. 13 of the systems with
high-resolution imaging are labeled as grade C lenses, meaning that
their nature is undetermined. The data for these systems, not shown in
this paper, come largely from WFPC2 snapshot observations. The
signal-to-noise ratio of these WFPC2 images is low compared to images
taken with ACS or WFC3 despite the longer exposure times. Most of the
remaining grade C systems are targets observed with NIR photometry and
adaptive optics, which proved not to be a very useful technique for the
follow-up of our candidates.

Ground-based data can be used in some cases to construct lens models and
measure precise Einstein radii: 9 out of 23 lenses with only CFHT
photometry are grade A lenses. 
The uncertainty on $R_{\mathrm{Ein}}$ for those lenses is still dominated by the $3\%$ systematic error, meaning that ground based photometry can sometimes be as good as space based imaging for the purpose of measuring Einstein radii.
For most systems however the
information is not enough to draw definite conclusions on their nature,
and in a few cases the data does not offer enough constraints to measure
Einstein radii, mostly because of the difficulty in detecting and
exploiting the counterimage as seen from the ground.
The range in Einstein radii covered by the grade A lenses in our sample is $5-15$ kpc, typically larger than those of other surveys such as SLACS, probing the mass in regions where the contribution of dark matter is larger.

Stellar masses of lens galaxies can be measured from ground-based data. 
Measurements of $M_*$ are robust to the inclusion of NIR data. NIR
should give more reliable stellar masses, since the blue background
sources contribute very little to the infrared flux. Our result suggests
that our measurements of the optical photometry of our lenses have
little contamination from the background sources, and that we
effectively deblended lens and source light.
Stellar masses of SL2S lenses cover the range $10^{11}-10^{12}M_\odot$, corresponding to massive ETGs.

In Paper IV we use all these measurements to put constraints on the
mass profile of massive early-type galaxies and its
evolution in the redshfit range $0.1 < z < 0.8$.

%-------------------------------------------------------------------------------

\acknowledgments

%\TODO{SHS}{Check SL2S grant support.}
% Boilerplate:
We thank our friends of the SLACS and SL2S collaborations for many
useful and insightful discussions over the course of the past years.
We thank V.N.~Bennert and M.~Bradac for their help in our observational campaign.
TT thanks S.W.~Allen and B.~Poggianti for
useful discussions.
RG acknowledges support from the Centre National des Etudes Spatiales
(CNES).
PJM acknowledges support from the Royal
Society in the form of a research fellowship. 
TT acknowledges support from the NSF through CAREER award NSF-0642621, and from
the Packard Foundation through a Packard Research Fellowship.
This research is based on XSHOOTER observations made with ESO Telescopes at the Paranal
Observatory under programme IDs 086.B-0407(A) and 089.B-0057(A).
This research is based on observations obtained with MegaPrime/MegaCam, a joint project of CFHT
and CEA/DAPNIA, and with WIRCam, a joint project of CFHT, Taiwan,
Korea, Canada and France, at the Canada-France-Hawaii Telescope (CFHT) which is operated
by the National Research Council (NRC) of Canada, the Institut National des
Sciences de l'Univers of the Centre National de la Recherche Scientifique
(CNRS) of France, and the University of Hawaii. This work is based in part on
data products produced at TERAPIX and the Canadian Astronomy Data Centre.
The authors would like to thank S. Arnouts, L. Van waerbeke and G. Morrison for giving access to the WIRCam data collected in W1 and W4 as part of additional CFHT programs. We are particularly thankful to Terapix for the data reduction of this dataset.
This research is supported by NASA through Hubble Space Telescope programs
GO-10876, GO-11289, GO-11588 and in part by the National Science Foundation
under Grant No. PHY99-07949, and is based on observations made with the
NASA/ESA Hubble Space Telescope and obtained at the Space Telescope Science
Institute, which is operated by the Association of Universities for Research in
Astronomy, Inc., under NASA contract NAS 5-26555, and at the W.M. Keck
Observatory, which is operated as a scientific partnership among the California
Institute of Technology, the University of California and the National
Aeronautics and Space Administration. The Observatory was made possible by the
generous financial support of the W.M. Keck Foundation. The authors wish to
recognize and acknowledge the very significant cultural role and reverence that
the summit of Mauna Kea has always had within the indigenous Hawaiian
community.  We are most fortunate to have the opportunity to conduct
observations from this mountain.

%-------------------------------------------------------------------------------

\bibliographystyle{apj}
\bibliography{references}

%-------------------------------------------------------------------------------

\end{document}